\documentclass[12pt]{iopart}
\usepackage{graphicx}
\begin{document}

\title{Dynamics of mobile interacting ferromagnetic films: theory and numerical implementation}

\author{Andrea Benassi}
\address{Empa, Swiss Federal Laboratories for Materials Science and Technology, CH-8600 D\"{u}bendorf, Switzerland.}
\ead{andrea.benassi@empa.ch}

\begin{abstract}
Coating two nearby bodies with thin ferromagnetic films one obtains, below the Curie temperature, two interacting sets of magnetic domains. The dynamical properties of the bodies in presence of this domain interaction have never been investigated so far.
In this work I derive a set of equations to simultaneously describe both the domain evolution within the two films and the dynamics of the coated bodies. 
The shape, size and mobility of the domains can be easily controlled with an external magnetic field or properly choosing the material properties, thus unravelling how the
domain characteristics influence the system dynamics. This can be thus of great technological relevance, providing new means to control and actuate mechanical motion at the micro- and nano-scale.
\end{abstract}


\maketitle

\begin{figure}
\centering
\includegraphics[height=7cm,angle=0]{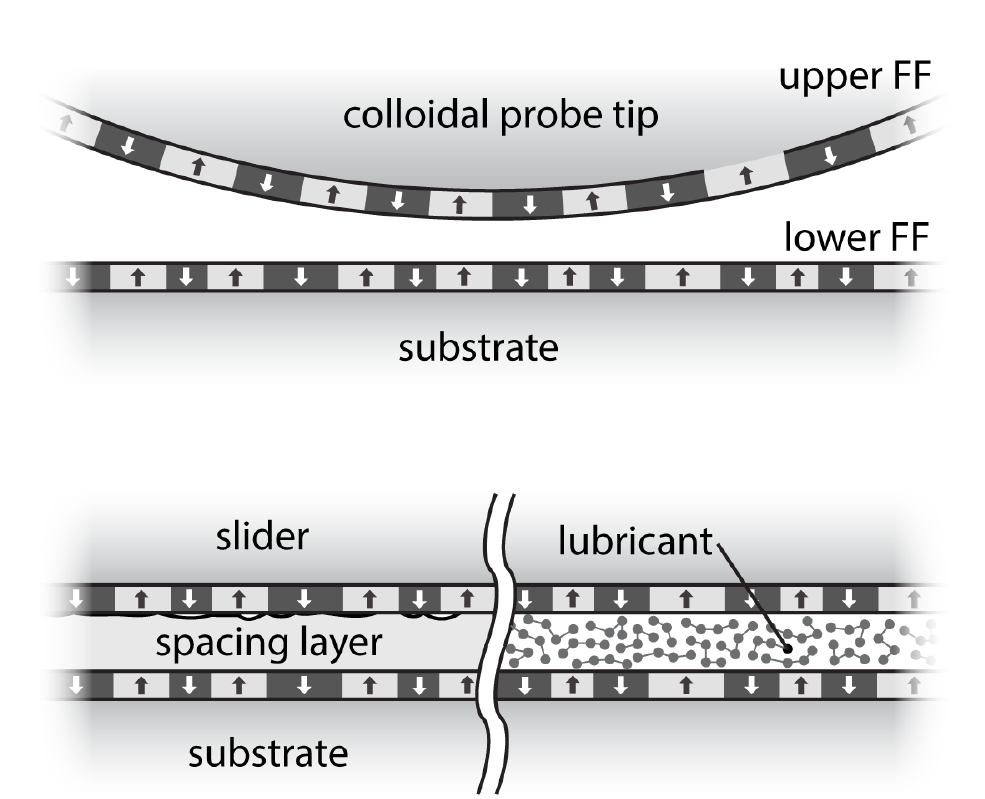}
\caption{Sketch of possible practical implementations of the film-on-film geometry. The upper panel shows a substrate and a colloidal probe tip both coated with ferromagnetic films, 
the relative sliding motion occurs in non-contact mode. The lower panel shows two possible setups for a contact mode sliding: in the left picture the two films are kept at constant 
distance by a non-magnetic spacing layer, in the right picture the two bodies are sliding in boundary lubrication regime, i.e. they are separated by few ordered molecular layers of lubricant.}
\label{figure1}
\end{figure}

The possibility to control friction, and thus the sliding motion of two bodies, has been extensively investigated both at the fundamental and 
applied level, being closely tied to progress in transportation, manufacturing, and energy conversion, and thus impacting on innumerable aspects of our health and environment.
Not all the control techniques available at the macro-scale, such as the use of lubricants, the surface patterning or the application of mechanical vibrations, are straightforwardly 
applicable to micro- and nano-mechanical systems because of the different scaling of physical laws with the system size.
At the micro and nano-scale however, new physical phenomena can be exploited for the control of motion, such as the atomic lattice commensurability or the superlubric transition. 
The possibility to control sliding friction through the occurrence of a structural phase transition in one or both the sliding bodies as been recently suggested \cite{mio}. This technique
allows to control the phase transition, an thus the sliding motion, dynamically and reversibly by means of an external electric field or by applying a pressure to the sliding bodies. 
Along the same lines, I believe that the sliding motion can be controlled coating the two sliding bodies with thin ferromagnetic films (FFs), as depicted in figure \ref{figure1}.
Below the Curie temperature, the presence of magnetic domains can in fact give rise to very strong forces able to modify the sliding dynamics, moreover the domain shape and size can 
be controlled by an external field, thus enabling for a dynamical and reversible control of motion.
The aim of this work is to develop the necessary mathematical and computational tools to investigate the mutual influence of magnetic domain interaction and sliding motion of the 
coated bodies, namely  to set up a system of equations to simultaneously describe the domain dynamics within each FF and the sliding motion of the coated bodies.\\
Motivated by the data storage technology needs, the physics of magnetic domains in thin ferromagnetic films has been deeply and extensively investigated in the past decades.
This work focuses on FFs with perpendicular anisotropy, i.e. the easy axis of the magnetization is perpendicular to the film surface, this behavior is typical of Co/Pt and Fe/Ni 
multilayers, permalloy and garnet films to name a few. 
In these FFs, the domains exhibit stable disordered maze-like patterns but, under the influence of an external magnetic field the domains can be ordered into parallel 
stripes or bubble lattices \cite{bertotti,hubert}. The characteristic domain size, ranging from tens of nm to tens of $\mu$m, can be controlled by the materials and the film thickness 
\cite{baltz} while 
changing the deposition rate one controls the homogeneity of the FFs, promoting the presence of defects and impurities that serve as pinning sites for the domains, 
thus controlling the domain mobility \cite{pierce}.\\ 
Experiments to test this new suggested control mechanism can be set up in several ways.
As illustrated in figure \ref{figure1}, the two FFs can be grown on a substrate and on a colloidal probe tip having a large curvature radius so that their interaction can be probed with an 
atomic force microscopy apparatus in non-contact mode. The atomic force microscope can also be used to study the contact sliding between two large plates \cite{tang}, a 
meso-scale friction tester \cite{wang} or a surface force apparatus \cite{carpick} can be used as well.
When the two coated bodies slide in contact mode the two FFs can be protected from wear by a capping layer and they can be kept at constant distance with sub-nanometric precision by a non-magnetic spacing layer. Mechanical friction can be reduced by the use of lubricants.\\
In section \ref{singolo} we recall, generalize and comment the existing theory for the description of the domain evolution in a single isolated FF; in section \ref{duefilm} we extend the 
theory in order to
describe the case of two interacting parallel films; in section \ref{eppursimuove} we introduce the Newton equations to describe the FF dynamics, i.e. the coated bodies motion; finally 
in section \ref{imp} we discuss the algorithms for the numerical implementation 
of the new set of equations.

\section{Single Ferromagnetic Film}
\label{singolo}
%------------------------------------------------------
%------------------------------------------------------
The magnetic properties of a FF below its Curie temperature can be modeled in several ways. Traditional modeling encompasses statistical 
approaches, like the Preisach one, as well as micromagnetics \cite{bertotti,hubert}. While the former allow to describe the hysteresis loop of 
macroscopic samples without any clue on the real microscopic domain dynamics, the latter can be used to access the fine details of the domain structure and motion, although the 
higher computational cost limits the size of the treatable systems. An intermediate phase-field approach exists which, starting from the micromagnetics equations, by means of suitable 
approximations, allow to investigate the detailed domain dynamics in FFs over large length-scales such as the ones accessible to Magnetic Force Microscopy 
(MFM), few $\mu$m$^2$, or to optical techniques, hundreds of $\mu$m$^2$.
The first numerical works using this approach have been performed by E. Jagla who investigated the possible stable and unstable domain patterns in thin perpendicular anisotropy 
FFs \cite{jagla1} and the role of the domain dynamics in determining the hysteresis loop shape \cite{jagla2}. In more recent works the same kind of modeling has been adopted to 
investigate return point memory effects \cite{pierce}, Barkhausen avalanche distributions and critical exponents \cite{benassimag1}, and the role of defects in the domain 
reorientation under the influence of an oscillating external field \cite{kudo}. Recently we have also demonstrated that this kind of modeling reproduces quantitatively both the domain 
dynamics at the micro-scale and the macroscopic hysteresis properties of exchange-bias Co/Pt multilayers \cite{benassimag2}. 

\subsection{Hamiltonian and domain equation of motion}
\begin{figure}
\centering
\includegraphics[height=10cm,angle=0]{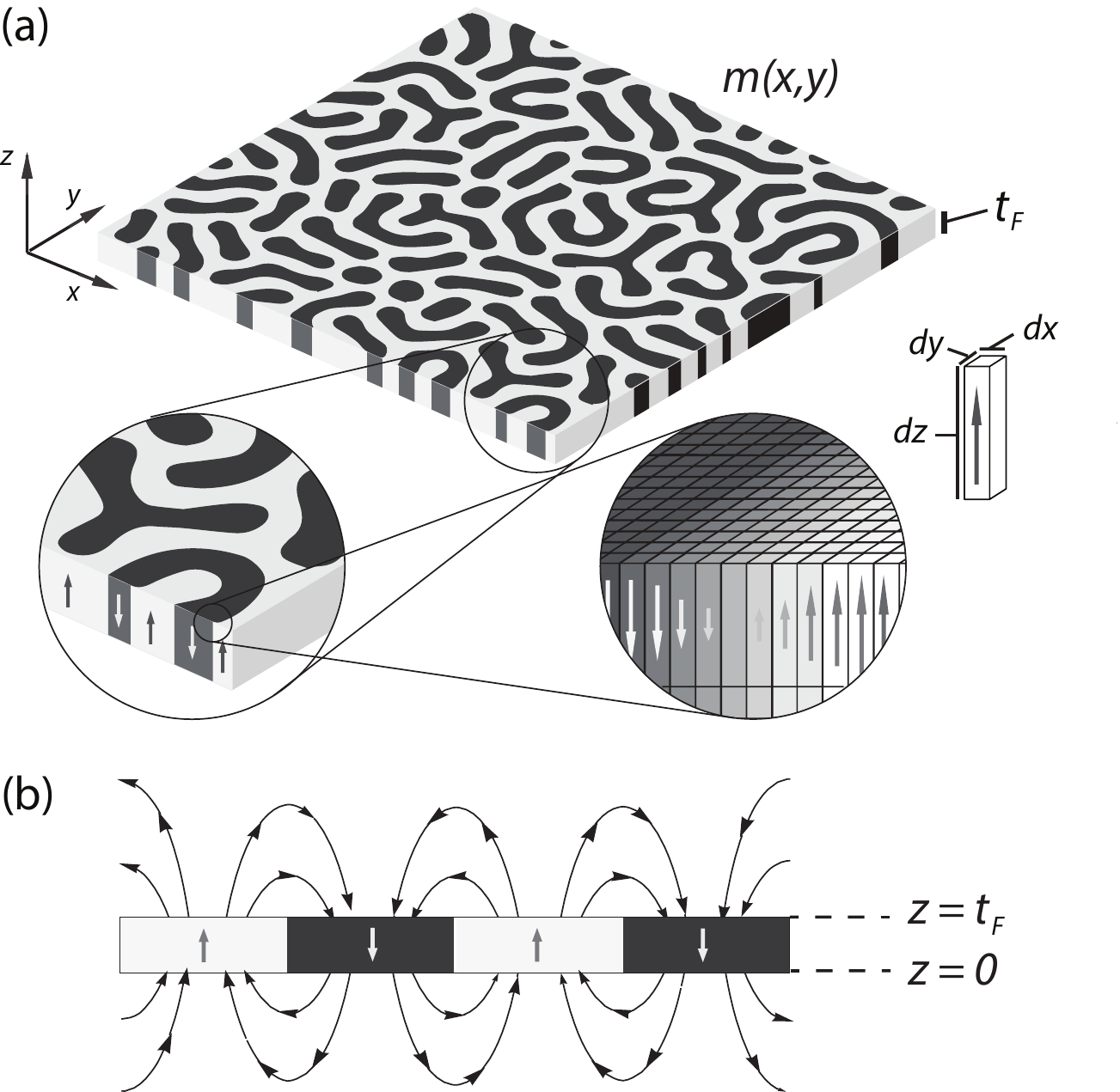}
\caption{(a) Sketch of the ferromagnetic film with perpendicular anisotropy, the zooms show the upward and downward oriented domains and the structure of a domain wall in our scalar approximation. (b) Film cross section and stray field stream lines.}
\label{figure2}
\end{figure}

The starting point for the study of the domain dynamics is the Landau-Lifshitz-Gilbert equation (LLGE), ruling the precession 
motion of the magnetization vector $\mathbf{M}(\mathbf{r},t)$, associated to the infinitesimal medium volume $d^3\mathbf{r}$, around a local field $\mathbf{B}(\mathbf{r},t)$ due to 
the presence of the rest of the medium and to external sources:
\begin{equation}
\frac{\partial \mathbf{M}(\mathbf{r},t)}{\partial t}=-\gamma \mathbf{M}(\mathbf{r},t) \times \Bigg( \mathbf{B}(\mathbf{r},t) -\eta \frac{\partial \mathbf{M}(\mathbf{r},t)}{\partial t} \Bigg),
\label{llg}
\end{equation} 
where $\gamma$ is the giromagnetic ratio of the electron spin and $\eta$ is a characteristic damping time of the material, representing the irreversible energy transfer to 
microscopic degrees of freedom such as magnons, phonons, and eddy-currents. The magnetization can be written as $\mathbf{M}(\mathbf{r},t)=M_s \mathbf{m}(\mathbf{r},t)$, 
separating its modulus, i.e. the saturation magnetization $M_s$, a material parameter,  from the dimensionless versor $\mathbf{m}$. Defining the dimensionless constant 
$\alpha=\gamma \eta M_s$, in the limit $\alpha\ll 1$, (\ref{llg}) can be rewritten as \cite{gilbert}:
\begin{equation}
\frac{\partial \mathbf{M}(\mathbf{r},t)}{\partial t}=-\gamma \mathbf{M}(\mathbf{r},t) \times  \mathbf{B}(\mathbf{r},t) - \frac{\gamma \alpha}{M_s} \mathbf{M}(\mathbf{r},t) 
\times  \mathbf{M}(\mathbf{r},t) \times  \mathbf{B}(\mathbf{r},t). 
\label{llg2}
\end{equation}
Theoretical calculations and experimental measurements have demonstrated that the assumption $\alpha\ll 1$ is fulfilled by most of the ferromagnetic materials in their bulk, multilayer and thin film forms, although in the latter case $\alpha$ can be slightly dependent on film thickness and growing conditions \cite{gilmore,mizukami1,mizukami2,barman}.   
The field $\mathbf{B}$ depends on the material properties and on the sample shape and it is known once the system Hamiltonian $\mathcal{H}$ is given:
 \begin{equation}
\mathbf{B}(\mathbf{r},t)=-\frac{1}{M_s}\frac{\delta \mathcal{H}[\mathbf{m}(\mathbf{r},t)]}{\delta \mathbf{m}(\mathbf{r},t)}+\mathbf{Q}(\mathbf{r},t).
\label{funder}
\end{equation}
The first term is the functional derivative of the Hamiltonian while the second one is a Gaussian stochastic process representing the thermal fluctuations the system experiences being 
in contact with an heat-bath at temperature $T$ \cite{brown}. More precisely we can characterize the stochastic process $\mathbf{Q}$ giving its average and correlation:
\begin{equation}
\langle \mathbf{Q}(\mathbf{r},t) \rangle=0 \qquad  \langle \mathbf{Q}(\mathbf{r},t) \mathbf{Q}(\mathbf{r}',t')\rangle=2 K_B T \frac{\alpha}{\gamma M_s}\delta(t-t')
\delta(\mathbf{r}-\mathbf{r}'),
\end{equation}
$K_B$ is the Boltzmann constant, from the two Dirac delta in the correlation function we see that the stochastic process is uncorrelated in time and space.
At finite temperature (\ref{llg2}) can be seen as a Langevin equation, in which the balancing of the competing damping and stochastic terms allows to sample the precession trajectories 
from a canonical ensemble.\\
Micromagnetic simulations can be performed starting from (\ref{llg2}) and evolving the magnetization in time on a three dimensional mesh \cite{miltat}, the field $\mathbf{B}$ felt by 
every magnetic dipole, associated to the infinitesimal medium volume, will be the sum of the filed due to all the other dipoles. This non locality, together with the full vectorial 
treatment of the problem, is responsible for the high computational cost of this kind of simulations limiting the size of the simulated samples.
However, to describe the physics of certain systems with a specific symmetry, one component of the magnetization might be more relevant than the others. This is the case of 
perpendicular anisotropy FF in which, except for the domain wall regions, the magnetization is mostly aligned perpendicular to the film plane as depicted in figure \ref{figure2}(a).
In this simplified picture the magnetization is assumed to be uniform along the $z$ axis, in the approximation of thin domain walls \cite{jagla2}, only its $z$ component plays a
relevant role, thus the domain dynamics can be solely described by a scalar function $m$ varying on the film plane only, i.e. $\mathbf{M}\equiv M_s m(x,y)\hat{\mathbf{z}}=M_s m(\mathbf{r}_\parallel)\hat{\mathbf{z}}$. Notice 
that, by construction, $m(x,y)$ must vary continuously in the interval $[-1,+1]$. 
%------------------------------------------------------
%------------------------------------------------------
To give the magnetization a preferential orientation (easy-axis) along the $z$ direction, the Hamiltonian must contain a quadratic term in $m$:
\begin{equation}
\mathcal{H}_{anisotropy}=-\frac{K_u}{2} \int m(\mathbf{r}_\parallel)^2\; d^3\mathbf{r},
\end{equation}
so that the energy is lowered the most when $\vert m \vert \rightarrow 1$ irrespective of the sign, i.e. irrespective of the upward or downward orientation of the magnetic dipole 
moments. $K_u$ is the anisotropy constant of the material determining the strength of the energy gain with the dipole moments alignment.\\ 
Every dipole moment of the film feels a stray field (also referred to as demagnetizing or 
dipolar field) due to the presence of the other dipole moments, as illustrated in figure \ref{figure2}(b). For our simple geometry of a plane FF of thickness $t_F$, the stray field energy can be calculated exactly starting 
from the potential $\phi$ due to a magnetization distribution:
\begin{equation}
\phi(\mathbf{r},t)=\frac{M_s}{4 \pi}\Bigg( -\int \frac{\nabla\cdot \mathbf{m}(\mathbf{r}',t)}{\vert \mathbf{r}-\mathbf{r}'\vert}d^3\mathbf{r}'
+\oint \frac{\mathbf{m}(\mathbf{r}',t) \cdot \hat{\mathbf{n}}}{\vert \mathbf{r}-\mathbf{r}'\vert}d\Sigma'
\Bigg),
\end{equation}
where $\hat{\mathbf{n}}$ is the versor normal to the surface, the first integral is on the film volume, the second one is on the film surface. 
With our choice for $\mathbf{m}(\mathbf{r},t)$ only the surface integral survives and the potential reduces to: 
\begin{equation}
\phi(\mathbf{r},t)=\frac{M_s}{4 \pi}\int\Bigg(\frac{m(\mathbf{r}'_\parallel,t)}{\sqrt{(\mathbf{r}_{\parallel}-\mathbf{r}'_{\parallel})^2+(z-t_F)^2}}
-\frac{m(\mathbf{r}'_\parallel,t)}{\sqrt{(\mathbf{r}_{\parallel}-\mathbf{r}'_{\parallel})^2+z^2}}
\Bigg)d^2\mathbf{r}'_{\parallel},
\label{sega}
\end{equation}
now $\mathbf{r}_{\parallel}$ and $\mathbf{r}'_{\parallel}$ span the $xy$ plane only, i.e. $(\mathbf{r}_{\parallel}-\mathbf{r}'_{\parallel})^2=(x-x')^2+(y-y')^2$ . The first contribution to the integral comes from the upper surface ($z'=t_F$), the 
second one comes from the lower surface of 
the film ($z'=0$), see figure \ref{figure2} (b). The self-energy of a magnetization distribution can be calculated as:
 \begin{equation}
 \eqalign{
\mathcal{H}_{stray}&=\frac{\mu_0 M_s}{2}\int \nabla \phi(\mathbf{r},t) \cdot \mathbf{m}(\mathbf{r},t) \; d^3\mathbf{r}\cr
&=\frac{\mu_0 M_s}{2} \bigg(\int\phi(\mathbf{r},t)\;m(\mathbf{r})  
\;d^2\mathbf{r}_\parallel \bigg)\bigg\vert^{z=t_F}_{z=0},
}
\label{minchia}
\end{equation}
the factor $1/2$ is to avoid the double counting in the sum of all the dipole-dipole contribution (i.e. the double integral on $\mathbf{r}$ and $\mathbf{r}'$), $\mu_0$ is the vacuum 
permeability. The second step
comes from a simple integration by parts, taking into account that the magnetization is directed along $z$ only and does not vary along the film thickness, i.e. is not a function of $z$.
We can write explicitly the energy substituting (\ref{sega}) into (\ref{minchia}):
\begin{equation}
\mathcal{H}_{stray}=\frac{\mu_0 M_s^2}{4 \pi}\int \Bigg(\frac{m(\mathbf{r}'_\parallel,t) m(\mathbf{r}_\parallel,t)}{\vert\mathbf{r}_{\parallel}-\mathbf{r}'_{\parallel}\vert}-\frac{m(\mathbf{r}'_\parallel,t) m(\mathbf{r}_\parallel,t)}{\sqrt{(\mathbf{r}_{\parallel}-\mathbf{r}'_{\parallel})^2+t_F^2}}
\Bigg)d^2\mathbf{r}_{\parallel}\;d^2\mathbf{r}'_{\parallel}.
\label{orco2}
\end{equation}
Having in mind the streamlines of the magnetic field generated by a single dipole moment, it is easy to understand that, in order to minimize the total energy, each dipole tries to align 
oppositely the neighboring ones.\\ 
Due to electronic interactions the system gains energy leaving the neighboring 
dipoles aligned along the same direction. The Hamiltonian term accounting for this behavior must contain a space derivative of $\mathbf{m}(\mathbf{r},t)$ in order to lose energy at every spatial variation 
of the magnetization:
\begin{equation}
\mathcal{H}_{exchange}=\frac{A}{2} \int \big[ \nabla \mathbf{m}(\mathbf{r},t)\big]^2\; d^3\mathbf{r},
\end{equation}
where $A$ is the exchange stiffness, representing the energy cost to misalign neighboring dipole moments, and the square is necessary to treat in the same way upward and 
downward spatial variations. This term is in competition with the stray field one and the characteristic domain size arise from the balancing of the two, see section \ref{pheno}.
In order for our model to be able to describe the domain manipulation via external magnetic field $\mathbf{H}_{ext}$, the last ingredient we need to include in the Hamiltonian is given by:
 \begin{equation}
\mathcal{H}_{extern}=-\mu_0 M_s\int \mathbf{H}_{ext} \cdot \mathbf{m}(\mathbf{r},t) \; d^3\mathbf{r}=-\mu_0 M_s\int H_{ext} \; m(\mathbf{r},t) \; d^3\mathbf{r},
\label{esterno}
\end{equation}
the second step comes from the assumption that the external field is completely aligned along the $z$ axis. 
Notice the absence of the $1/2$ factor with respect to (\ref{minchia}), this is in fact the energy contribution due to an external field, not a self-energy. 
To summarize, the full Hamiltonian for a single FF reads:
\begin{equation}
\eqalign{
\mathcal{H}&=\int \Bigg[ -K_u\frac{m(\mathbf{r}_\parallel)^2}{2}+\frac{A}{2}[\nabla \mathbf{m}(\mathbf{r},t)]^2 -\mu_0 M_s \; m(\mathbf{r}_\parallel) \; H_{ext} \cr
&+\frac{\mu_0 M_s^2}{4 \pi t_F}
\int d^2\mathbf{r}'_{\parallel}\Bigg(\frac{m(\mathbf{r}_\parallel) m(\mathbf{r}'_\parallel)}{\vert\mathbf{r}_{\parallel}-\mathbf{r}'_{\parallel}\vert}-\frac{m(\mathbf{r}_\parallel) m(\mathbf{r}'_\parallel)}{\sqrt{(\mathbf{r}_{\parallel}-\mathbf{r}'_{\parallel})^2+t_F^2}}
\Bigg)\Bigg]d^3\mathbf{r},
 }
\label{full}
\end{equation}
the $1/t_F$ in the stray field term comes from the need of restoring a volume integral in (\ref{orco2}), being the integrand independent of $z$, we can simply put $d^2\mathbf{r}=d^2\mathbf{r}dz/t_F=d^3\mathbf{r}/t_F$.
From the functional derivative (\ref{funder}) we can thus calculate the field $\mathbf{B}$ which is parallel to the $z$ axis (from now on we drop the subscript $\parallel$ and $\mathbf{r}$ and $\mathbf{r}'$  are intended to run on the $xy$ plane only):
\begin{equation}
\eqalign{
\mathbf{B}&=\Bigg[\frac{K_u}{M_s}\; m(\mathbf{r}) +\mu_0 \; H_{ext}
-\frac{\mu_0 M_s}{2 \pi t_F }
\int \Bigg(\frac{m(\mathbf{r}')}{\vert\mathbf{r}-\mathbf{r}'\vert}\cr
&-\frac{m(\mathbf{r}')}{\sqrt{(\mathbf{r}-\mathbf{r}')^2+t_F^2}}
\Bigg) d^2\mathbf{r}'\Bigg]\hat{\textbf{z}}
+Q(\mathbf{r},t)\hat{\textbf{z}}+\frac{A}{M_s}\;\nabla^2 \mathbf{m}(\mathbf{r},t),
}
\label{campo}
\end{equation}
the gradient term has been treated with the ``thin domain wall" approximation as described in \cite{jagla2}.
Substituting the previous expression into (\ref{llg2}) we immediately see that the fist term on the r.h.s. vanishes and we remain with:
\begin{equation}
\eqalign{
\frac{\partial m}{\partial t}&=\gamma \alpha\; \Bigg\{(1-m^2)
\Bigg[
\frac{K_u}{M_s}\; m+\mu_0 \; H_{ext} 
-\frac{\mu_0 M_s}{2 \pi t_F}\int \Bigg(\frac{m(\mathbf{r}')}{\vert\mathbf{r}-\mathbf{r}'\vert}\cr
&-\frac{m(\mathbf{r}')}{\sqrt{(\mathbf{r}-\mathbf{r}')^2+t_F^2}}
\Bigg)d^2\mathbf{r}'+Q(\mathbf{r},t)
\Bigg]
+\frac{A}{M_s}\;\nabla^2 m
\Bigg\}.
}
\label{orco}
\end{equation}

\subsection{Small thickness approximation and useful limits}
In the early works by E. Jagla the stray field term has been treated in the small thickness approximation $t_F\rightarrow 0$. If we power expand the second term in the r.h.s of 
(\ref{orco2}) for small $t_F$, we see that the zero order contribution cancels out with the first term leaving only the contribution in $t_F^2$ (the first order contribution is zero for parity reasons):
\begin{equation}
\mathcal{H}_{stray}=\frac{\mu_0 M_s^2 t_F^2}{8 \pi}\int \frac{m(\mathbf{r}',t) m(\mathbf{r},t)}{\vert\mathbf{r}-\mathbf{r}'\vert^3}
d^2\mathbf{r}\;d^2\mathbf{r}'.
\label{piccolit}
\end{equation}
From this simplified expression the tendency of the stray field to anti-align the dipole moments is immediately clear: if the spin at the point $\mathbf{r}$ is oriented in the same 
direction of the one at $\mathbf{r}'$ the product $m(\mathbf{r}') m(\mathbf{r})$ is positive and the total energy increase, to gain energy the two spin at $\mathbf{r}$ and $\mathbf{r}'$ 
must be oppositely oriented so that $\mathcal{H}_{stray}<0$.\\ 
It is also important to notice that, when completely saturated, i.e. $m(x,y)\equiv \pm 1\; \forall \; x,y$, the FF behaves like a uniformly charged plane capacitor. This means that the outer 
field is zero while the inner one is constant and it depends only on the material parameters. The field expression (\ref{campo}), with $m(\mathbf{r}',t)=1$, at zero temperature and in absence of any external field becomes:
\begin{equation}
B(\mathbf{r})=\frac{K_u}{M_s}-\frac{\mu_0 M_s}{2 \pi t_F}
\int \Bigg(\frac{1}{\vert\mathbf{r}-\mathbf{r}'\vert}-\frac{1}{\sqrt{(\mathbf{r}-\mathbf{r}')^2+t_F^2}}
\Bigg)d^2\mathbf{r}',
\label{invtra}
\end{equation}
the integrand depends only on $\vert\mathbf{r}-\mathbf{r}'\vert$, if the film is infinitely extended along $x$ and $y$, we have translational invariance, i.e. the integral over 
$\mathbf{r'}$ gives the same result for every $\mathbf{r}$. We can exploit this symmetry to solve the integral for the convenient choice $\mathbf{r}=0$:
\begin{equation}
\eqalign{
B&=\frac{K_u}{M_s}-\frac{\mu_0 M_s}{2 \pi t_F}
\int \Bigg(\frac{1}{\vert\mathbf{r}'\vert}-\frac{1}{\sqrt{\mathbf{r}'^2+t_F^2}}
\Bigg)d^2\mathbf{r}'=\cr
&=\frac{K_u}{M_s}-\frac{\mu_0 M_s}{2 \pi t_F}
\lim_{\ell \rightarrow \infty}\int_0^{\ell} r' dr'\int_0^{2\pi} \Bigg(\frac{1}{\vert\mathbf{r}'\vert}-\frac{1}{\sqrt{\mathbf{r}'^2+t_F^2}}
\Bigg)d\theta=\cr
&=\frac{K_u}{M_s}-\lim_{\ell \rightarrow \infty} \frac{\mu_0 M_s}{t_F}\big( 
\ell+t_F-\sqrt{\ell^2+t_F^2}
\big)=\frac{K_u}{M_s}-\mu_0 M_s,
}
\end{equation}
the second step has been obtained moving to polar coordinates. The same can be done for the energy density, i.e. the integrand 
in the r.h.s. of (\ref{full}) which, for $m(\mathbf{r}) m(\mathbf{r}')=1$ and calculated in the infinitesimal volume $d^3\mathbf{r}$ centered at $\mathbf{r}=0$, reads
\begin{equation}
\frac{\mathcal{H}}{d^3\mathbf{r}}=-\frac{K_u}{2}+\frac{\mu_0 M_s^2}{4 \pi t_F}
\int \Bigg(\frac{1}{\vert\mathbf{r}'\vert}-\frac{1}{\sqrt{\mathbf{r}'^2+t_F^2}}\Bigg)d^2\mathbf{r}'=-\frac{K_u}{2}+\frac{\mu_0 M_s^2}{2},
\end{equation}
again the last step holds for small $t_F$.

\subsection{Pinning disorder}
Under the influence of an external magnetic field $H_{ext}$, the magnetization of a ferromagnetic material can be manipulated, promoting nucleation, growth and 
coalescence of domains. However the magnetization does not vary smoothly with the external field strength, its dynamics is characterized by sudden jumps due to the discontinuous 
motion of the domain walls pinned by crystalline defects and grain boundaries, these jumps are known as Barkhausen avalanches. The disorder and the inhomogeneities of the material 
serve also as nucleation points at the initial stage of the magnetization reversal process.
The pinning effect due to the sample inhomogeneities can be included in our model by letting one ore more material properties fluctuate randomly on the $xy$ plane. Contrary to 
the disorder introduced by thermal fluctuations which changes in time, this new source of randomness is fixed once and for all at $t=0$.
The fluctuation can be introduced in the anisotropy constant (random anisotropy model), in the exchange stiffness constant (random bond model) or simply by means of a 
magnetic field $H_{random}$ (random field model):
\begin{equation}
\eqalign{
K_u(\mathbf{r})= K_u \; [1-c\; p(\mathbf{r})],\cr
A(\mathbf{r})= A \; [1-c\; p(\mathbf{r})],\cr
H_{random}(\mathbf{r})=c\; p(\mathbf{r}),
}
\label{rumore}
\end{equation}
here $c$ is a parameter determining the amplitude of the fluctuations, i.e. the strength of the pinning inhomogeneities, $K_u$ and $A$ are the 
macroscopic average material parameter, and $p(\mathbf{r})$ is a Gaussian 
stochastic process with $\langle p(\mathbf{r}) \rangle=0$ and a given correlation $\langle p(\mathbf{r}) p(\mathbf{r}')\rangle$.\\
As we will see in section \ref{imp}, the LLGE must be solved numerically on a discrete mesh of spacing $\Delta$, if this quantity is bigger than the characteristic length scale of
the FF inhomogeneities, the pinning disorder fluctuations will be uncorrelated (white noise), i.e. $\langle p(\mathbf{r}) p(\mathbf{r}')\rangle=\delta 
(\mathbf{r}-\mathbf{r}')$. More generally, the disorder can be correlated on a characteristic length scale dictated by the micro-structure of the FF, for instance the average crystalline 
grain size. A correlated random field has been recently used to study how the domain dynamics is affected by the presence of uncompensated spins, at the interface between 
ferromagnetic and anti-ferromagnetic films, in exchange-bias systems \cite{benassimag2}.\\
The physics of the three fluctuating noises is of course different: the fluctuations entering the Hamiltonian through the anisotropy or the exchange terms, which are quadratic in 
$m$, are not sensitive 
to the magnetization sign, i.e. they serve as nucleation points for both upward and downward oriented domains, and the up-to-down and down-to-up hysteresis semi-loops are 
exactly identical (with the same domain patterns). To have different nucleation points in the two hysteresis branches, we need to use a random field model, its 
Hamiltonian term is linear in $m$ and thus sensitive to the magnetization sign.
Typically, in a real FF, all the three sources of randomness are simultaneously present.

\subsection{Model parameters and domain behavior}
\label{pheno}
%------------------------------------------------------
\begin{figure}
\centering
\includegraphics[height=3.2cm,angle=0]{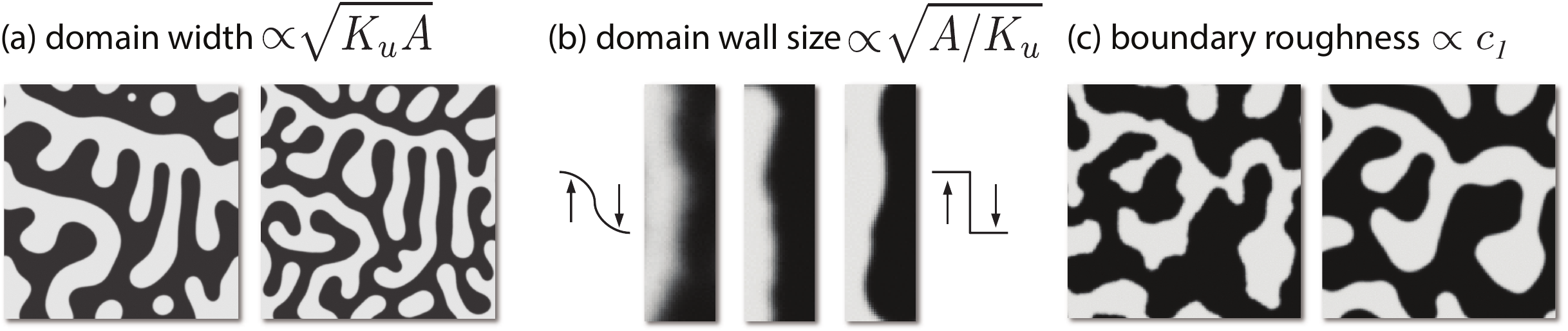}
\caption{Simulated domain morphologies. (a) Two different stable domain configurations obtained decreasing $\sqrt{K_u A}$.
 (b) Domain walls of decreasing thickness obtained decreasing $\sqrt{A/K_u}$. (c) The same stable domain configuration obtained decreasing the anisotropy fluctuation strength $c$.}
\label{figure3}
\end{figure}
%------------------------------------------------------
The material parameters $K_u$, $A$ and $c$ determine the domain morphology and dynamics.
Starting from a simplified version of (\ref{full}), it is easy to demonstrate that the energy cost of a domain wall is proportional to $\sqrt{K_u A}$ \cite{bertotti,hubert,landau}, this is also easy to 
understand qualitatively: in a domain wall the magnetic dipoles are misaligned with respect to the neighbors, thus we must pay an energy proportional to $A$, but they are also 
misaligned with respect to the anisotropy easy axis, thus we pay also an anisotropy energy proportional to $K_u$. 
The characteristic average domain width is set by the competition of the stray field and the energy cost of a wall,  while the former tends to create a large amount of small domains, 
to demagnetize the film,  the latter tries to minimize the number of domain walls and thus of domains. As explicitly visible from (\ref{orco2}), the stray field strength depends solely on 
the film thickness, thus at fixed $t_F$, the domain size can be tuned varying the domain wall cost only and it goes as $\sqrt{K_u A}$. On the other hand, working
with a given material, we can tune the domain width by choosing properly the film thickness, the thickness dependence is however non-trivial \cite{baltz,kaplan}. 
Figure \ref{figure3} (a) shows the magnetization $m(x,y)$ resulting from two different simulations with the same thickness but different $K_u A$ product.
Again starting from (\ref{full}) it is easy to show that the domain wall thickness goes as $\sqrt{A/K_u}$, figure \ref{figure3} (b) shows how the simulated domain walls get narrower as 
we decrease the $A/K_u$ ratio.  
For many ferromagnetic materials the $K_u$ and $A$ values are tabulated \cite{bertotti}, for thin films and multilayers they are known to be dependent on the thickness and 
deposition conditions. However, $K_u$ and $A$ can also be estimated starting from a measured domain image, in fact, knowing both the real domain width and domain wall thickness 
we fix both the product $K_u A$ and the ration $A/K_u$, and a unique couple of values for $K_u$ and $A$ which satisfies both the conditions exists. Using this idea we can extract
the $K_u$ and $A$ values directly from the simulations once the simulated domain morphology resembles the measured one \cite{benassimag2}. For Co/Pt multilayers 
we found a very good agreement between the calculated and experimentally estimated $K_u$ values whereas for $A$ we always found an overestimation. This last result is easily explainable  
recalling that in (\ref{orco}) we neglected a term in $(\nabla m)^2$, for the sake of simplicity and computational feasibility, assuming  that its main effect is simply to renormalize the $A$ constant \cite{jagla2}.
The inhomogeneity strength $c$ is also important in determining the domain shape, as shown in figure \ref{figure3} (c): increasing the inhomogeneity of the FF, the domain 
boundaries become more irregular. The $c$ value also determines the domain mobility under the influence of an external magnetic field, a large $c$ value results is a strong domain 
pinning with a very irregular and sudden domain motion, smaller $c$ values lead to a more smooth and continuous motion. For this reason the inhomogeneity strength can be 
estimated from the size distribution of the measured Barkhausen avalanches \cite{benassimag1}. 
The two remaining material parameters $M_s$ and $\alpha$ play no role in determining the domain characteristics, they simply define the absolute strength of the field and the absolute time scale for the domain motion respectively.
This is explicitly shown in section \ref{us} where (\ref{orco}) is rescaled in order to be dimensionless, $M_s$ becomes the unit field while $\alpha$ enters the unit time, ans both of them disappear from the rescaled equation.
\section{Two interacting ferromagnetic films}
\label{duefilm}
\begin{figure}
\centering
\includegraphics[height=5cm,angle=0]{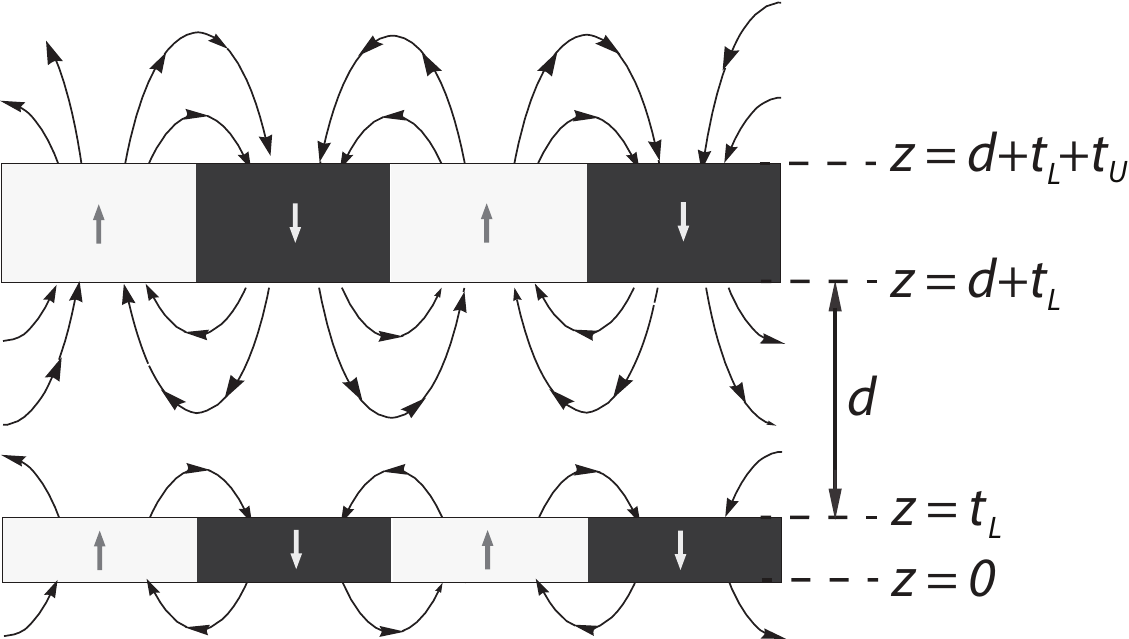}
\caption{Cross section of the two interacting films and stream lines of their field.}
\label{figure4}
\end{figure}

The possible experimental setups sketched in figure \ref{figure1} can be modeled  by the idealized geometry of two plane interacting FFs of thickness $t_U$ and $t_L$ kept at a given distance $d$, as sketched in figure \ref{figure4}.
%------------------------------------------------------
%------------------------------------------------------
\subsection{Hamiltonian and domain equation of motion}
When $d \rightarrow \infty$ the two films are well described by two independent Hamiltonians like (\ref{full}) where we simply 
replace $m$, $M_s$, $K_u$ and $A$ with $m_U$, $M_U$, $K_U$ and $A_U$ for the upper film and with $m_L$, $M_L$, $K_L$ and $A_L$ for the lower film. The full Hamiltonian is thus $\mathcal{H}=\mathcal{H}_U+\mathcal{H}_L$ with:
\begin{equation}
\eqalign{
\mathcal{H}_U&=\int_U \Bigg[ -K_U\frac{m_U(\mathbf{r})^2}{2}+\frac{A_U}{2}[\nabla m_U(\mathbf{r})]^2 -\mu_0 M_U \; m_U(\mathbf{r}) \; H_{ext} \cr
&+\frac{\mu_0 M_U^2}{4 \pi t_U}
\int\Bigg(\frac{m_U(\mathbf{r}) m_U(\mathbf{r}')}{\vert\mathbf{r}-\mathbf{r}'\vert}-\frac{m_U(\mathbf{r}) m_U(\mathbf{r}')}{\sqrt{(\mathbf{r}-\mathbf{r}')^2+t_U^2}}d^2\mathbf{r}'
\Bigg)\Bigg]d^3\mathbf{r},
 }
 \label{hu}
\label{fullupper}
\end{equation}
\begin{equation}
\eqalign{
\mathcal{H}_L&=\int_L\Bigg[ -K_L\frac{m_L(\mathbf{r})^2}{2}+\frac{A_L}{2}[\nabla m_L(\mathbf{r})]^2 -\mu_0 M_L \; m_L(\mathbf{r}) \; H_{ext} \cr
&+\frac{\mu_0 M_L^2}{4 \pi t_L}
\int\Bigg(\frac{m_L(\mathbf{r}) m_L(\mathbf{r}')}{\vert\mathbf{r}-\mathbf{r}'\vert}-\frac{m_L(\mathbf{r}) m_L(\mathbf{r}')}{\sqrt{(\mathbf{r}-\mathbf{r}')^2+t_L^2}}d^2\mathbf{r}'
\Bigg)\Bigg] d^3\mathbf{r}.
 }
\label{hl}
\label{fulllower}
\end{equation}
When the two films are brought closer, each one feels the field generated by the other and their magnetic domains start to interact. The new Hamiltonian term responsible for this interaction can be obtained calculating the energy of the upper film in presence of
the field generated by the lower one, to this aim we start from (\ref{esterno}), where the integral is on the upper film volume, and we transform it into a surface integral exactly as we did for (\ref{minchia}):
\begin{equation}
\eqalign{
\mathcal{H}_{int}&=\mu_0 M_U\int_U \nabla \phi_L(\mathbf{r},t) \cdot \mathbf{m}(\mathbf{r},t) \; d^3\mathbf{r}\cr
&=\mu_0 M_U\int_U \frac{\partial\phi_L(\mathbf{r},t)}{\partial z} \; m_U(\mathbf{r},t) \; d^3\mathbf{r}\cr
&=\mu_0 M_U\bigg(\int \phi_L(\mathbf{r},t) \; m_U(\mathbf{r},t) d^2\mathbf{r} \bigg)\bigg\vert^{z=t_L+d+t_U}_{z=t_L+d},
}
\end{equation}
notice again the absence of the factor $1/2$, this is not a self-energy term. Substituting the potential due to the lower film $\phi_L(\mathbf{r},t)$, given by (\ref{sega}), we obtain:
\begin{equation}
\eqalign{
\mathcal{H}_{int}&=\frac{\mu_0 M_U M_L}{4 \pi}\int \Bigg(-
\frac{m_U(\mathbf{r},t) m_L(\mathbf{r}',t)}{\sqrt{(\mathbf{r}-\mathbf{r}')^2+d^2}}\cr
&+\frac{m_U(\mathbf{r},t) m_L(\mathbf{r}',t)}{\sqrt{(\mathbf{r}-\mathbf{r}')^2+(d+t_U)^2}}+
\frac{m_U(\mathbf{r},t) m_L(\mathbf{r}',t)}{\sqrt{(\mathbf{r}-\mathbf{r}')^2+(d+t_L)^2}}\cr
&-\frac{m_U(\mathbf{r},t) m_L(\mathbf{r}',t)}{\sqrt{(\mathbf{r}-\mathbf{r}')^2+(d+t_U+t_L)^2}}
\Bigg)d^2 \mathbf{r}\; d^2\mathbf{r}',
}
\label{intera}
\end{equation}
a term which is symmetric with respect to the upper and lower films and depends on their material and geometric parameters. Notice 
that the same expression can be derived calculating the interaction energy of the lower film in the field due to the upper one, and we should use the expression:
\begin{equation}
\mathcal{H}_{int}=\mu_0 M_U\bigg(\int \phi_U(\mathbf{r},t) \; m_L(\mathbf{r},t) d^2\mathbf{r} \bigg)\bigg\vert^{z=t_L}_{z=0},
\end{equation}
and calculate $ \phi_U(\mathbf{r},t)$ analogously to (\ref{sega}). The two field expressions obtained differentiating $\mathcal{H}_{int}$ with respect to $m_L$ or $m_U$ become:
\begin{equation}
\eqalign{
B_{U}&=-\frac{\mu_0 M_L}{4 \pi\; t_U}\int \Bigg(-
\frac{m_L(\mathbf{r}',t)}{\sqrt{(\mathbf{r}-\mathbf{r}')^2+d^2}}-\frac{m_L(\mathbf{r}',t)}{\sqrt{(\mathbf{r}-\mathbf{r}')^2+(d+t_U+t_L)^2}}\cr
&+\frac{m_L(\mathbf{r}',t)}{\sqrt{(\mathbf{r}-\mathbf{r}')^2+(d+t_U)^2}}+
\frac{m_L(\mathbf{r}',t)}{\sqrt{(\mathbf{r}-\mathbf{r}')^2+(d+t_L)^2}}
\Bigg)d^2 \mathbf{r}',
}
\label{fu}
\end{equation}
\begin{equation}
\eqalign{
B_{L}&=-\frac{\mu_0 M_U }{4 \pi\; t_L}\int \Bigg(-
\frac{m_U(\mathbf{r}',t)}{\sqrt{(\mathbf{r}-\mathbf{r}')^2+d^2}}-\frac{m_U(\mathbf{r}',t)}{\sqrt{(\mathbf{r}-\mathbf{r}')^2+(d+t_U+t_L)^2}}\cr
&+\frac{m_U(\mathbf{r}',t)}{\sqrt{(\mathbf{r}-\mathbf{r}')^2+(d+t_U)^2}}+
\frac{m_U(\mathbf{r}',t)}{\sqrt{(\mathbf{r}-\mathbf{r}')^2+(d+t_L)^2}}
\Bigg)d^2 \mathbf{r}',
}
\label{fl}
\end{equation}
with the field acting on one film depending only on the magnetization of the other one. Also here, to perform the functional derivative of $\mathcal{H}=\mathcal{H}_U+\mathcal{H}_L+\mathcal{H}_{int}$, we have to restore a volume integral in (\ref{intera}), before differentiating with respect to $m_U$ we put $d^2\mathbf{r}=d^3\mathbf{r}/t_U$
and before differentiating with respect to $m_L$ we put $d^2\mathbf{r}'=d^3\mathbf{r}'/t_L$.\\
The two magnetizations $m_L$ and $m_U$ evolve in time according to (\ref{llg2}), and we end up with the two coupled equations:
\begin{equation}
\eqalign{
\frac{\partial m_U}{\partial t}&=\gamma \alpha_U\; \Bigg\{(1-m_U^2)
\Bigg[
\frac{K_U}{M_U}\; m_U+\mu_0 \; H_{ext}
-\frac{\mu_0 M_U}{2 \pi t_U}\int \Bigg(\frac{m_U(\mathbf{r}',t)}{\vert\mathbf{r}-\mathbf{r}'\vert}\cr
&-\frac{m_U(\mathbf{r}',t)}{\sqrt{(\mathbf{r}-\mathbf{r}')^2+t_U^2}}
\Bigg)d^2\mathbf{r}'
-\frac{\mu_0 M_L}{4 \pi\; t_U}\int \Bigg(-
\frac{m_L(\mathbf{r}',t)}{\sqrt{(\mathbf{r}-\mathbf{r}')^2+d^2}}\cr
&-\frac{m_L(\mathbf{r}',t)}{\sqrt{(\mathbf{r}-\mathbf{r}')^2+(d+t_U+t_L)^2}}
+\frac{m_L(\mathbf{r}',t)}{\sqrt{(\mathbf{r}-\mathbf{r}')^2+(d+t_U)^2}}\cr
&+\frac{m_L(\mathbf{r}',t)}{\sqrt{(\mathbf{r}-\mathbf{r}')^2+(d+t_L)^2}}
\Bigg)d^2 \mathbf{r}'
+Q_U(\mathbf{r},t)\Bigg]
+\frac{A_U}{M_U}\;\nabla^2 m_U
\Bigg\},
}
\label{vau}
\end{equation}
\begin{equation}
\eqalign{
\frac{\partial m_L}{\partial t}&=\gamma \alpha_L\; \Bigg\{(1-m_L^2)
\Bigg[
\frac{K_L}{M_L}\; m_L+\mu_0 \; H_{ext}
-\frac{\mu_0 M_L}{2 \pi t_L}\int \Bigg(\frac{m_L(\mathbf{r}',t)}{\vert\mathbf{r}-\mathbf{r}'\vert}\cr
&-\frac{m_L(\mathbf{r}',t)}{\sqrt{(\mathbf{r}-\mathbf{r}')^2+t_L^2}}
\Bigg)d^2\mathbf{r}'
-\frac{\mu_0 M_U}{4 \pi\; t_L}\int \Bigg(-
\frac{m_U(\mathbf{r}',t)}{\sqrt{(\mathbf{r}-\mathbf{r}')^2+d^2}}\cr
&-\frac{m_U(\mathbf{r}',t)}{\sqrt{(\mathbf{r}-\mathbf{r}')^2+(d+t_U+t_L)^2}}
+\frac{m_U(\mathbf{r}',t)}{\sqrt{(\mathbf{r}-\mathbf{r}')^2+(d+t_U)^2}}\cr
&+\frac{m_U(\mathbf{r}',t)}{\sqrt{(\mathbf{r}-\mathbf{r}')^2+(d+t_L)^2}}
\Bigg)d^2 \mathbf{r}'
+Q_L(\mathbf{r},t)\Bigg]
+\frac{A_L}{M_L}\;\nabla^2 m_L
\Bigg\},
}
\label{val}
\end{equation}
with the thermal noise properties:
\begin{equation}
\eqalign{
\langle Q_U(\mathbf{r},t) \rangle=0 \quad  \langle Q_U(\mathbf{r},t) Q_U(\mathbf{r}',t')\rangle=2 K_B T \frac{\alpha_U}{\gamma M_U}\delta(t-t') 
\delta(\mathbf{r}-\mathbf{r}'),\cr
\langle Q_L(\mathbf{r},t) \rangle=0 \quad  \langle Q_L(\mathbf{r},t) Q_L(\mathbf{r}',t')\rangle=2 K_B T \frac{\alpha_L}{\gamma M_L}\delta(t-t') 
\delta(\mathbf{r}-\mathbf{r}').
}
\end{equation}

\subsection{Small thickness approximation and useful limits}
\label{emo}
Notice that our interaction energy expression we can evaluate it in some simple limiting cases.
Very useful, to this aim, is the limit in which two films of the same material ($M_U=M_L=M_s$) and thickness ($ t_U=t_L=t_F$) are brought in close contact ($d=0$), i.e. we obtain a 
single film of thickness $2 t_F$.
In fact, being the films identical their magnetization must behave in the same way, with $m_U=m_L=m$ the stray field energy terms contained in (\ref{hu}) and (\ref{hl}) cancel out with part of the interaction term and we are left with:
\begin{equation}
\eqalign{
\mathcal{H}_{int}+\mathcal{H}_{stray}&=\frac{\mu_0 M_s^2}{4 \pi}\int \Bigg(\frac{m(\mathbf{r}',t) m(\mathbf{r},t)}{\vert\mathbf{r}-\mathbf{r}'\vert}\cr
&-\frac{m(\mathbf{r}',t) m(\mathbf{r},t)}{\sqrt{(\mathbf{r}-\mathbf{r}')^2+(2 t_F)^2}}
\Bigg)d^2\mathbf{r}\;d^2\mathbf{r}',
}
\label{attaccati}
\end{equation}
which is exactly (\ref{orco2}) for a film with with thickness $2 t_F$ $q.e.d.$.
In practice, when $d\rightarrow 0$, the magnetization of the two films is exactly the same even if, as discussed in section \ref{pheno}, an isolated film with larger thickness should display smaller domains having a larger stray field.
This happens because the film-film interaction compensates the stray field energy difference of the isolated films, an analytical demonstration is possible in the small thickness limit $t_L,t_U\rightarrow 0$ where we can write:
\begin{equation}
\eqalign{
\mathcal{H}_{stray}+\mathcal{H}_{int}=
\frac{\mu_0 M_U^2 t_U^2}{8 \pi}\int \frac{m_U(\mathbf{r}',t) m_U(\mathbf{r},t)}{\vert\mathbf{r}-\mathbf{r}'\vert^3}
d^2\mathbf{r}\;d^2\mathbf{r}'\cr
+\frac{\mu_0 M_L^2 t_L^2}{8 \pi}\int \frac{m_L(\mathbf{r}',t) m_L(\mathbf{r},t)}{\vert\mathbf{r}-\mathbf{r}'\vert^3}
d^2\mathbf{r}\;d^2\mathbf{r}'\cr
+\frac{\mu_0 M_U M_L}{4 \pi}t_L\;t_U\int
\frac{m_U(\mathbf{r},t) m_L(\mathbf{r}',t)\big[(\mathbf{r}-\mathbf{r}')^2-2d^2\big]}{\big[(\mathbf{r}-\mathbf{r}')^2+d^2\big]^{5/2}}
\;d^2 \mathbf{r}\;d^2 \mathbf{r}',
}
\end{equation}
and from the functional derivatives we get the fields:
\begin{equation}
\eqalign{
B_U&=
-\frac{\mu_0 M_U t_U}{4 \pi}\int \frac{m_U(\mathbf{r}',t)}{\vert\mathbf{r}-\mathbf{r}'\vert^3}
d^2\mathbf{r}'\cr
&-\frac{\mu_0 M_L t_L}{4 \pi} \int
\frac{m_L(\mathbf{r}',t)\big[(\mathbf{r}-\mathbf{r}')^2-2d^2\big]}{\big[(\mathbf{r}-\mathbf{r}')^2+d^2\big]^{5/2}}
d^2 \mathbf{r}',
}
\end{equation}
\begin{equation}
\eqalign{
B_L&=
-\frac{\mu_0 M_L t_L}{4 \pi}\int \frac{m_L(\mathbf{r}',t)}{\vert\mathbf{r}-\mathbf{r}'\vert^3}
d^2\mathbf{r}'\cr
&-\frac{\mu_0 M_U t_U}{4 \pi}\int
\frac{m_U(\mathbf{r},t)\big[(\mathbf{r}-\mathbf{r}')^2-2d^2\big]}{\big[(\mathbf{r}-\mathbf{r}')^2+d^2\big]^{5/2}}
d^2 \mathbf{r}',
}
\end{equation}
notice now that for $d=0$ the two expressions become identical, the total field felt by the two films is thus the same and, even if  $t_U\neq t_L$, their domains must behave in the same way and display the same patterns, i.e. $m_U=m_L$.
This finding is in agreement with the experimentally observed domain behavior in the limit of $d \rightarrow 0$ \cite{baltz}.\\
We can perform another important consistency check of our equations in the saturation limit $m_L(x,y)=m_U(x,y)\equiv\pm1 \; \forall \; x,y$: as previously discussed, when a FF is uniformly magnetized its outer field must vanish. 
In this limit the expression for the outer field generated by the lower film (\ref{fu}) becomes:
\begin{equation}
\eqalign{
B_U&=-\frac{\mu_0 M_L}{4 \pi \;t_U}\int \Bigg(-
\frac{1}{\sqrt{(\mathbf{r}-\mathbf{r}')^2+d^2}}
+\frac{1}{\sqrt{(\mathbf{r}-\mathbf{r}')^2+(d+t_U)^2}}\cr
&+\frac{1}{\sqrt{(\mathbf{r}-\mathbf{r}')^2+(d+t_L)^2}}
-\frac{1}{\sqrt{(\mathbf{r}-\mathbf{r}')^2+(d+t_U+t_L)^2}}
\Bigg)d^2 \mathbf{r}',
}
\end{equation}
like in (\ref{invtra}) the translational invariance allows us to calculate the field in the convenient point $\mathbf{r}=0$, in polar coordinates we have:
\begin{equation}
\eqalign{
B_U&=-\frac{\mu_0 M_L}{2t_U}\lim_{\ell \rightarrow\infty}\int_{0}^{\ell} \Bigg(-
\frac{\mathbf{r}'}{\sqrt{\mathbf{r}'^2+d^2}}
+\frac{\mathbf{r}'}{\sqrt{\mathbf{r}'^2+(d+t_U)^2}}\cr
&+\frac{\mathbf{r}'}{\sqrt{\mathbf{r}'^2+(d+t_L)^2}}
-\frac{\mathbf{r}'}{\sqrt{\mathbf{r}'^2+(d+t_U+t_L)^2}}
\Bigg)d^2 \mathbf{r}'\cr
&=-\frac{\mu_0 M_L}{2t_U}\lim_{\ell \rightarrow \infty}\Big(-\sqrt{\ell^2+d^2}+
\sqrt{\ell^2+(d+t_U)^2}\cr
&+\sqrt{\ell^2+(d+t_L)^2}
-\sqrt{\ell^2+(d+t_U+t_L)^2}
\Big)=0\qquad q.e.d.
}
\label{secco}
\end{equation}

\section{Mobile interacting films}
\label{eppursimuove}

\subsection{Force calculation}
%------------------------------------------------------
\begin{figure}
\centering
\includegraphics[height=3.5cm,angle=0]{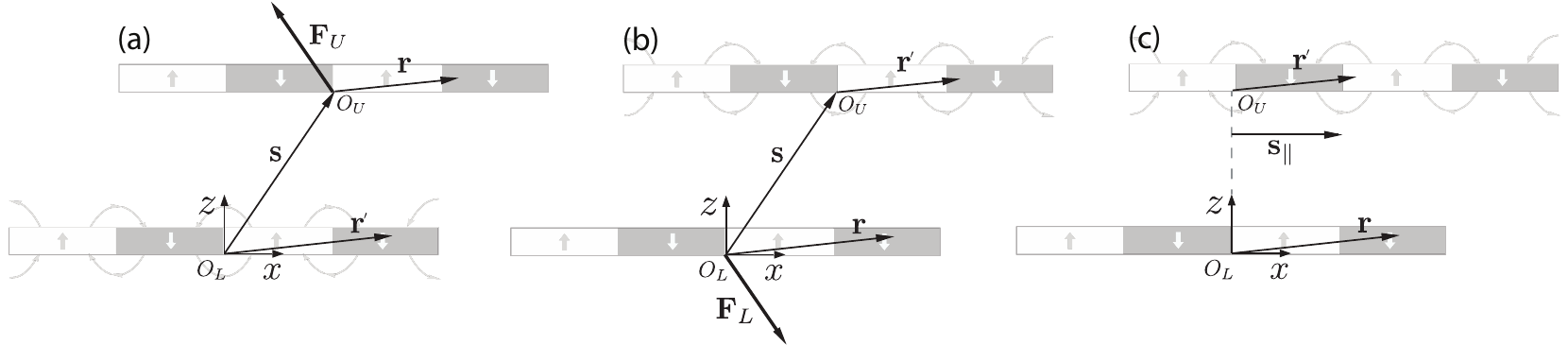}
\caption{Sketch of the reference frame adopted and the vectors involved in the calculation of (a) $\mathbf{F}_U$ and (b) $\mathbf{F}_L$. In this two cases the origin of the upper frame $O_U$, moves forward together with the upper film. (c) is the equivalent of (b) in the opposite 
picture in which $O_U$ is immobile and always aligned with $O_L$, and the magnetization is shifted forward with respect to $O_U$.}
\label{figure5}
\end{figure}
%------------------------------------------------------
The equations derived so far describe the domain evolution into the two FFs, now we want to study the dynamics of the two FF, i.e. the motion of the two coated bodies. To this aim we need to calculate the magnetic force that each film exerts on the 
other at every time $t$ given the magnetization distributions $m_U(\mathbf{r},t)$ and $m_L(\mathbf{r},t)$. At this point it is convenient to define a vector $\mathbf{s}(t)$ that connects a reference point in the lower film $O_L$
to a reference point in the upper film $O_U$. Looking at figure \ref{figure5} (a) it is easy to see that $\mathbf{s}(t)=[s_x(t),s_y(t),d(t)]=[\mathbf{s}_\parallel(t),d(t)]$ represents the relative displacement between the two films, notice that $O_L$ is also the 
center of our reference frame.
Due to the presence of the lower film, a field $\mathbf{B}_U$ exists in the upper film volume, and it exerts a force on each infinitesimal dipole moment, the total force on the upper film $\mathbf{F}_U$ is thus obtained integrating over all this infinitesimal contribution. 
Being the force on a single dipole moment proportional to $ (\mathbf{m}\cdot \nabla) \mathbf{B}$ (see Appendix A), we obtain:
\begin{equation}
\eqalign{
\mathbf{F}_U(t)&=M_U\int_U m_U(\mathbf{r},t) \frac{\partial \mathbf{B}_U[\mathbf{r}+\mathbf{s}_\parallel, m_L(t)]}{\partial z}\;d^3\mathbf{r}\cr
&=M_U \bigg(\int m_U(\mathbf{r},t) \mathbf{B}_U[\mathbf{r}+\mathbf{s}_\parallel, m_L(t)]\;d^2\mathbf{r}\bigg)\bigg\vert^{z=t_L+d+t_U}_{z=t_L+d},
}
\label{fump}
\end{equation}
where the field due to the lower film is calculated in the displaced position of the upper film $\mathbf{r}+\mathbf{s}_\parallel$ as depicted in figure \ref{figure5} (a), the last step comes again from a simple integration by parts. 
Notice also that, to study the domain evolution we only needed the $z$ component of $\mathbf{B}_U$, namely (\ref{fu}), now instead we need the full vector that can be calculated from the gradient of the potential (\ref{sega}):
\begin{equation}
\eqalign{
\mathbf{B}_U(\mathbf{r})&=\frac{\mu_0 M_L}{4 \pi}\int m_L(\mathbf{r}',t)\Bigg(\frac{\mathbf{r}-\mathbf{r}'+(z-t_L) \;\hat{\mathbf{z}}}{[(\mathbf{r}-\mathbf{r}')^2+(z-t_L)^2]^{3/2}}\cr
&-\frac{\mathbf{r}-\mathbf{r}'+z \;\hat{\mathbf{z}}}{[(\mathbf{r}-\mathbf{r}')^2+z^2]^{3/2}}
\Bigg)d^2 \mathbf{r}'.
}
\label{vettoriale}
\end{equation}
Substituting into (\ref{fump}) we get the following force:
\begin{equation}
\eqalign{
\mathbf{F}_{U}=-\frac{\mu_0 M_U M_L}{4 \pi}\int m_U(\mathbf{r},t) m_L(\mathbf{r}',t)\Bigg(
\frac{\mathbf{r}-\mathbf{r}'+\mathbf{s}_\parallel+d \;\hat{\mathbf{z}}}{[(\mathbf{r}-\mathbf{r}'+\mathbf{s}_\parallel)^2+d^2]^{3/2}}\cr
-\frac{\mathbf{r}-\mathbf{r}'+\mathbf{s}_\parallel+(d+t_U)\;\hat{\mathbf{z}}}{[(\mathbf{r}-\mathbf{r}'+\mathbf{s}_\parallel)^2+(d+t_U)^2]^{3/2}}-
\frac{\mathbf{r}-\mathbf{r}'+\mathbf{s}_\parallel+(d+t_L)\;\hat{\mathbf{z}}}{[(\mathbf{r}-\mathbf{r}'+\mathbf{s}_\parallel)^2+(d+t_L)^2]^{3/2}}\cr
+\frac{\mathbf{r}-\mathbf{r}'+\mathbf{s}_\parallel+(d+t_U+t_L)\;\hat{\mathbf{z}}}{[(\mathbf{r}-\mathbf{r}'+\mathbf{s}_\parallel)^2+(d+t_U+t_L)^2]^{3/2}}
\Bigg)d^2 \mathbf{r}\;d^2 \mathbf{r}'.
}
\label{delirio1}
\end{equation}
Conversely, to calculate the force acting on the lower film, we have to start from:
\begin{equation}
\mathbf{F}_L(t)=M_L \bigg(\int m_L(\mathbf{r},t) \mathbf{B}_L[\mathbf{r}-\mathbf{s}_\parallel, m_U(t)]\;d^2\mathbf{r}\bigg)\bigg\vert^{z=t_L}_{z=0},
\end{equation}
now the field felt by the lower film is given by:
\begin{equation}
\eqalign{
\mathbf{B}_L(\mathbf{r})=\frac{\mu_0 M_U}{4 \pi}\int m_U(\mathbf{r}',t)\Bigg(\frac{\mathbf{r}-\mathbf{r}'+(z-d-t_U-t_L) \;\hat{\mathbf{z}}}{[(\mathbf{r}-\mathbf{r}')^2+(z-d-t_U-t_L)^2]^{3/2}}\cr
-\frac{\mathbf{r}-\mathbf{r}'+(z-d-t_L) \;\hat{\mathbf{z}}}{[(\mathbf{r}-\mathbf{r}')^2+(z-d-t_L)^2]^{3/2}}
\Bigg)d^2 \mathbf{r}',
}
\end{equation}
and it must be calculated considering the upper film displaced with respect to the axes origin, i.e. in the points $\mathbf{r}'+\mathbf{s}_\parallel$, according to figure \ref{figure5} (b) 
Substituting into the force expression we get:
\begin{equation}
\eqalign{
\mathbf{F}_L=-\frac{\mu_0 M_U M_L}{4 \pi}\int m_U(\mathbf{r}',t) m_L(\mathbf{r},t)\Bigg(\frac{\mathbf{r}-\mathbf{r}'-\mathbf{s}_\parallel-d \;\hat{\mathbf{z}}}
{[(\mathbf{r}-\mathbf{r}'-\mathbf{s}_\parallel)^2+d^2]^{3/2}}\cr
-\frac{\mathbf{r}-\mathbf{r}'-\mathbf{s}_\parallel-(d+t_U) \;\hat{\mathbf{z}}}
{[(\mathbf{r}-\mathbf{r}'-\mathbf{s}_\parallel)^2+(d+t_U)^2]^{3/2}}
-\frac{\mathbf{r}-\mathbf{r}'-\mathbf{s}_\parallel-(d+t_L) \;\hat{\mathbf{z}}}
{[(\mathbf{r}-\mathbf{r}'-\mathbf{s}_\parallel)^2+(d+t_L)^2]^{3/2}}\cr
+\frac{\mathbf{r}-\mathbf{r}'-\mathbf{s}_\parallel-(d+t_U+t_L) \;\hat{\mathbf{z}}}
{[(\mathbf{r}-\mathbf{r}'-\mathbf{s}_\parallel)^2+(d+t_U+t_L)^2]^{3/2}}
\Bigg)d^2 \mathbf{r}\;d^2 \mathbf{r}',
}
\label{delirio2}
\end{equation}
renaming $\mathbf{r}$ into $\mathbf{r}'$ it is immediate to see that $\mathbf{F}_L=-\mathbf{F}_U$ as the Newton's third law requires.
Expressions (\ref{delirio1}) and (\ref{delirio2}) can be simplified in the limit of small thickness $t_U,t_L\rightarrow 0$:
\begin{equation}
\eqalign{
\mathbf{F}_{U}=\frac{3 \mu_0 M_U M_L}{4 \pi}\;t_U\;t_L\int \frac{m_U(\mathbf{r},t) m_L(\mathbf{r}',t)}{[(\mathbf{r}'-\mathbf{r}+\mathbf{s}_\parallel)^2+d^2]^{7/2}} \bigg[\bigg(\mathbf{r}'-\mathbf{r}+\mathbf{s}_\parallel\bigg)\cr
 \bigg(\vert \mathbf{r}'-\mathbf{r}+\mathbf{s}_\parallel\vert^2-4 d^2\bigg)+\hat{\mathbf{z}}\;d\;\bigg(\vert \mathbf{r}'-\mathbf{r}+\mathbf{s}_\parallel\vert^2-\frac{2}{3}d\bigg)
\bigg]d^2 \mathbf{r}'\;d^2 \mathbf{r}.
}
\end{equation}
In section \ref{emo} we demonstrated that the outer field of the FFs goes to zero when they saturate, for this reason the total force between them is expected to vanish as well. This can be easily proved putting $m_U=m_L\equiv\pm 1 \;\forall \; x,y$ in (\ref{delirio1}) or (\ref{delirio2}).\\
Before moving further it is necessary to stress that the LLGEs (\ref{vau}) and (\ref{val}) have been derived for two parallel FFs perfectly aligned, if we let the upper film to move we have to introduce the displacement vector $\mathbf{s}(t)$ in the calculation of the interaction term.
In the lower film equation we have to use the field $B_L(\mathbf{r})$ exerted by the upper one, we can still use (\ref{fl}) which is valid when $O_U$ is perfectly aligned with $O_L$, and account for the negative displacement of $O_L$ with respect to $O_U$ replacing $\mathbf{r}$ with 
$\mathbf{r}-\mathbf{s}_\parallel$: 
\begin{equation}
\eqalign{
\frac{\partial m_L}{\partial t}=\gamma \alpha_L\; \Bigg\{(1-m_L^2)
\Bigg[
\frac{K_L}{M_L}\; m_L+\mu_0 \; H_{ext}
-\frac{\mu_0 M_L}{2 \pi t_L}\int \Bigg(\frac{m_L(\mathbf{r}',t)}{\vert\mathbf{r}-\mathbf{r}'\vert}\cr
-\frac{m_L(\mathbf{r}',t)}{\sqrt{(\mathbf{r}-\mathbf{r}')^2+t_L^2}}
\Bigg)d^2\mathbf{r}'
-\frac{\mu_0 M_U}{4 \pi\; t_L}\int \Bigg(-
\frac{m_U(\mathbf{r}',t)}{\sqrt{(\mathbf{r}-\mathbf{r}'-\mathbf{s}_\parallel)^2+d^2}}\cr
-\frac{m_U(\mathbf{r}',t)}{\sqrt{(\mathbf{r}-\mathbf{r}'-\mathbf{s}_\parallel)^2+(d+t_U+t_L)^2}}
+\frac{m_U(\mathbf{r}',t)}{\sqrt{(\mathbf{r}-\mathbf{r}'-\mathbf{s}_\parallel)^2+(d+t_U)^2}}\cr
+\frac{m_U(\mathbf{r}',t)}{\sqrt{(\mathbf{r}-\mathbf{r}'-\mathbf{s}_\parallel)^2+(d+t_L)^2}}
\Bigg)d^2 \mathbf{r}'
+Q_L(\mathbf{r},t)\Bigg]
+\frac{A_L}{M_L}\;\nabla^2 m_L
\Bigg\}.
}
\label{vall}
\end{equation}
The same can be done for the upper film, considering the field (\ref{fu}) calculated in the position $\mathbf{r}+\mathbf{s}_\parallel$ to account for the forward shift of $O_U$ with respect to $O_L$:
\begin{equation}
\eqalign{
\frac{\partial m_U}{\partial t}=\gamma \alpha_U\; \Bigg\{(1-m_U^2)
\Bigg[
\frac{K_U}{M_U}\; m_U+\mu_0 \; H_{ext}
-\frac{\mu_0 M_U}{2 \pi t_U}\int \Bigg(\frac{m_U(\mathbf{r}',t)}{\vert\mathbf{r}-\mathbf{r}'\vert}\cr
-\frac{m_U(\mathbf{r}',t)}{\sqrt{(\mathbf{r}-\mathbf{r}')^2+t_U^2}}
\Bigg)d^2\mathbf{r}'
-\frac{\mu_0 M_L}{4 \pi\; t_U}\int \Bigg(-
\frac{m_L(\mathbf{r}',t)}{\sqrt{(\mathbf{r}-\mathbf{r}'+\mathbf{s}_\parallel)^2+d^2}}\cr
-\frac{m_L(\mathbf{r}',t)}{\sqrt{(\mathbf{r}-\mathbf{r}'+\mathbf{s}_\parallel)^2+(d+t_U+t_L)^2}}
+\frac{m_L(\mathbf{r}',t)}{\sqrt{(\mathbf{r}-\mathbf{r}'+\mathbf{s}_\parallel)^2+(d+t_U)^2}}\cr
+\frac{m_L(\mathbf{r}',t)}{\sqrt{(\mathbf{r}-\mathbf{r}'+\mathbf{s}_\parallel)^2+(d+t_L)^2}}
\Bigg)d^2 \mathbf{r}'
+Q_U(\mathbf{r},t)\Bigg]
+\frac{A_U}{M_U}\;\nabla^2 m_U
\Bigg\}.
}
\label{vauu}
\end{equation}
As we will see in section \ref{reci}, the properties of the Fourier transforms will allow us to transform the displacement in the interaction term into a shift of the magnetization, we will thus recover the old expressions (\ref{vau}) and (\ref{val}) but with a modified 
magnetization.

\subsection{Equation of motion}
As illustrated in figure \ref{figure1}, the possible practical setups to measure the magnetic interaction between the FFs consist of a rigid substrate and a mobile slider, we thus need a single equation of motion to evolve the displacement vector $\mathbf{s(t)}$ of the upper film, while the lower one is kept fixed.
In most of the sliding systems of interest for micro-mechanics and tribology, the slider can be driven at constant force $F_\parallel$ or at constant velocity, the latter case is typically modeled driving the slider through a spring which represents the elastic stiffness of the driving apparatus. For instance, an AFM tip is typically modeled by a spring $k_\parallel$ along the sliding direction, accounting for the 
torsional stiffness of the cantilever, and a spring $k_\perp$ perpendicular to the sliding plane, representing the vertical bending stiffness of the cantilever. 
When the slider and the substrate are kept in contact, a force $F_\perp$ can be added to load the slider and modify the contact properties, this force can also be added to effectively model the adhesion force between the two bodies.
From this considerations, a very general form of the equation for motion of the slider (upper film), is given by:
\begin{equation}
m \frac{\partial^2 \mathbf{s}(t)}{\partial t^2}=\mathbf{F}_U[\mathbf{s}(t)]+\mathbf{F}_{driving}-\zeta m \frac{\partial \mathbf{s}(t)}{\partial t},
\label{bega}
\end{equation}
where $m$ is the slider mass and $\zeta$ a damping coefficient. The driving force $\mathbf{F}_{driving}=(F_\parallel,0,F_\perp)$ has a component along the sliding direction $x$ and 
one perpendicular to it, along $z$. Analogously to the term $-\eta \partial \mathbf{M}/\partial t$ in (\ref{llg}), the viscous damping disposes off the energy with a characteristic time $1/\zeta$ 
representing the dissipation through microscopic mechanical degrees of freedom of slider and driving apparatus.
This last equation, coupled with (\ref{vau}) and (\ref{val}), completely describe the dynamics of the two sliding bodies and their magnetization.

\section{Numerical implementation}
\label{imp}

\subsection{Unit system}
\label{us}
For the numerical implementation of our set  of equations it is worth to choose a dimensionless unit system. The single film LLGE (\ref{orco}) can be rewritten in dimensionless units factorizing $\mu_0M_s$ in the r.h.s. and defining the film thickness $t_F$ as the unit length, so 
that $\tilde{\mathbf{r}}=\mathbf{r}/t_F$, and $1/\gamma\alpha\mu_0 M_s$ as a unit time, so that $\tilde{t}=t\; \gamma\alpha\mu_0 M_s$. With this substitution we finally 
get:
\begin{equation}
\eqalign{
\frac{\partial m}{\partial \tilde{t}}&=(1-m^2)
\Bigg[
a\; m+h_{ext} 
-\frac{1}{2 \pi }\int \Bigg(\frac{m(\tilde{\mathbf{r}}')}{\vert\tilde{\mathbf{r}}-\tilde{\mathbf{r}}'\vert}\cr
&-\frac{m(\tilde{\mathbf{r}}')}
{\sqrt{(\tilde{\mathbf{r}}-\tilde{\mathbf{r}}')^2+1}}
\Bigg)d^2\tilde{\mathbf{r}}'
+q(\tilde{\mathbf{r}},\tilde{t})
\Bigg]
+b\;\nabla^2 m,
}
\label{finale}
\end{equation}
where $a=K_u/\mu_0 M_s^2$ and $b=A/\mu_0 M_s^2 t_F^2$ are the dimensionless uniaxial anisotropy and exchange stiffness respectively, while $h_{ext}=H_{ext}/M_s$ 
and $q=Q/\mu_0 M_s$.  
In this dimensionless system the statistical properties of the thermal fluctuations become: 
 \begin{equation}
\langle q(\tilde{\mathbf{r}},\tilde{t}) \rangle=0 \qquad  \langle q(\tilde{\mathbf{r}},\tilde{t}) q(\tilde{\mathbf{r}}',\tilde{t}')\rangle=2 \widetilde{K_B T} \alpha^2 \delta(\tilde{t}-\tilde{t}') 
\delta(\tilde{\mathbf{r}}-\tilde{\mathbf{r}}'),
\end{equation}
with $\widetilde{K_B T }=K_B T/\mu_0 M_s^2 t_F^3$ dimensionless temperature.\\
We proceed in the same way for the coupled LLGEs (\ref{vall}) and (\ref{vauu}) ruling the domain dynamics in two interacting FF, in this case however, we have to choose one of the two 
films to be the reference, expressing all the fields in units of its saturation magnetization, and all the distances in units of its thickness. As a consequence of this choice  
the dimensionless equations become asymmetric. 
Choosing the lower film as a reference we can define:
\begin{equation}
\eqalign{
\tilde{t}=t\; \gamma \alpha_L \mu_0 M_L, \qquad \tilde{\mathbf{r}}=\frac{\mathbf{r}}{t_L}, \qquad h_{ext}=\frac{H_ext}{M_L},
\cr
a_L=\frac{K_L}{\mu_0 M_L^2}, \qquad b_L=\frac{A_L}{\mu_0 M_L^2 t_L^2}, \qquad q_L=\frac{Q_L}{\mu_0 M_L},\cr
a_U=\frac{K_U}{\mu_0 M_L M_U}, \qquad b_U=\frac{A_U}{\mu_0 M_L M_U t_L^2}, \qquad q_U=\frac{Q_U}{\mu_0 M_L},\cr  
\xi =\frac{M_U}{M_L}, \qquad \tilde{t}_U=\frac{t_U}{t_L}, \qquad \tilde{d}=\frac{d}{t_L}, \qquad \nu=\frac{\alpha_U}{\alpha_L}, 
}
\label{tabrutta}
\end{equation}
with these definitions we get:
\begin{equation}
\eqalign{
\frac{\partial m_U}{\partial \tilde{t}}=\nu\; \Bigg\{(1-m_U^2)
\Bigg[
a_U\; m_U+\frac{h_{ext}}{\xi}
-\frac{\xi}{2 \pi \;\tilde{t}_U}\int \Bigg(\frac{m_U(\tilde{\mathbf{r}}',\tilde{t})}{\vert\tilde{\mathbf{r}}-\tilde{\mathbf{r}}'\vert}\cr
-\frac{m_U(\tilde{\mathbf{r}}',\tilde{t})}{\sqrt{(\tilde{\mathbf{r}}-\tilde{\mathbf{r}}')^2+\tilde{t}_U^2}}
\Bigg)d^2\tilde{\mathbf{r}}'
-\frac{1}{4 \pi\; \tilde{t}_U}\int \Bigg(-
\frac{m_L(\tilde{\mathbf{r}}',\tilde{t})}{\sqrt{(\tilde{\mathbf{r}}-\tilde{\mathbf{r}}'+\tilde{\mathbf{s}}_\parallel)^2+\tilde{d}^2}}\cr
-\frac{m_L(\tilde{\mathbf{r}}',\tilde{t})}{\sqrt{(\tilde{\mathbf{r}}-\tilde{\mathbf{r}}'+\tilde{\mathbf{s}}_\parallel)^2+(\tilde{d}+\tilde{t}_U+1)^2}}
+\frac{m_L(\tilde{\mathbf{r}}',\tilde{t})}{\sqrt{(\tilde{\mathbf{r}}-\tilde{\mathbf{r}}'+\tilde{\mathbf{s}}_\parallel)^2+(\tilde{d}+\tilde{t}_U)^2}}\cr
+\frac{m_L(\tilde{\mathbf{r}}',\tilde{t})}{\sqrt{(\tilde{\mathbf{r}}-\tilde{\mathbf{r}}'+\tilde{\mathbf{s}}_\parallel)^2+(\tilde{d}+1)^2}}
\Bigg]d^2 \tilde{\mathbf{r}}'
+q_U(\tilde{\mathbf{r}},\tilde{t})\Bigg)
+b_U\;\nabla^2 m_U
\Bigg\},
}
\label{ultimateu}
\end{equation}
\begin{equation}
\eqalign{
\frac{\partial m_L}{\partial \tilde{t}}=(1-m_L^2)
\Bigg[
a_L\; m_L+ h_{ext}
-\frac{1}{2 \pi}\int \Bigg(\frac{m_L(\tilde{\mathbf{r}}',\tilde{t})}{\vert\tilde{\mathbf{r}}-\tilde{\mathbf{r}}'\vert}\cr
-\frac{m_L(\tilde{\mathbf{r}}',\tilde{t})}{\sqrt{(\tilde{\mathbf{r}}-\tilde{\mathbf{r}}')^2+1}}
\Bigg)d^2\tilde{\mathbf{r}}'
-\frac{\xi}{4 \pi}\int \Bigg(-
\frac{m_U(\tilde{\mathbf{r}}',\tilde{t})}{\sqrt{(\tilde{\mathbf{r}}-\tilde{\mathbf{r}}'-\tilde{\mathbf{s}}_\parallel)^2+\tilde{d}^2}}\cr
-\frac{m_U(\tilde{\mathbf{r}}',\tilde{t})}{\sqrt{(\tilde{\mathbf{r}}-\tilde{\mathbf{r}}'-\tilde{\mathbf{s}}_\parallel)^2+(\tilde{d}+\tilde{t}_U+1)^2}}
+\frac{m_U(\tilde{\mathbf{r}}',\tilde{t})}{\sqrt{(\tilde{\mathbf{r}}-\tilde{\mathbf{r}}'-\tilde{\mathbf{s}}_\parallel)^2+(\tilde{d}+\tilde{t}_U)^2}}\cr
+\frac{m_U(\tilde{\mathbf{r}}',\tilde{t})}{\sqrt{(\tilde{\mathbf{r}}-\tilde{\mathbf{r}}'-\tilde{\mathbf{s}}_\parallel)^2+(\tilde{d}+1)^2}}
\Bigg)d^2 \tilde{\mathbf{r}}'
+q_L(\tilde{\mathbf{r}},\tilde{t})\Bigg]
+b_L\;\nabla^2 m_L.
}
\label{ultimatel}
\end{equation}
In the dimensionless unit system the thermal noise becomes:
\begin{equation}
\eqalign{
\langle q_U(\tilde{\mathbf{r}},\tilde{t}) \rangle=0 \quad  \langle q_U(\tilde{\mathbf{r}},\tilde{t}) q_U(\tilde{\mathbf{r}}',\tilde{t}')\rangle=2 \widetilde{K_B T} \frac{\alpha_U\alpha_L}{\xi}\delta(\tilde{t}-\tilde{t}') 
\delta(\tilde{\mathbf{r}}-\tilde{\mathbf{r}}'),\cr
\langle q_L(\tilde{\mathbf{r}},\tilde{t}) \rangle=0 \quad  \langle q_L(\tilde{\mathbf{r}},\tilde{t}) q_L(\tilde{\mathbf{r}}',\tilde{t}')\rangle=2 \widetilde{K_B T} \alpha_L^2\delta(\tilde{t}-\tilde{t}') 
\delta(\tilde{\mathbf{r}}-\tilde{\mathbf{r}}'),
}
\end{equation}
with $\widetilde{K_B T}=K_B T/\mu_0 M_L^2 t_L^3$.\\
To conclude we need to put the Newton equation for the slider motion (\ref{bega}) in the same unit system defined by (\ref{tabrutta}):
\begin{equation}
\eqalign{
\widetilde{m} \frac{\partial^2 \tilde{\mathbf{s}}(t)}{\partial \tilde{t}^2}=-\frac{\xi}{4 \pi}\int m_U(\tilde{\mathbf{r}},t) m_L(\tilde{\mathbf{r}}',t)\Bigg(
\frac{\tilde{\mathbf{r}}'-\tilde{\mathbf{r}}+\tilde{\mathbf{s}}_\parallel+\tilde{d} \;\hat{\mathbf{z}}}{[(\tilde{\mathbf{r}}'-\tilde{\mathbf{r}}+\tilde{\mathbf{s}}_\parallel)^2+\tilde{d}^2]^{3/2}}\cr
-\frac{\tilde{\mathbf{r}}'-\tilde{\mathbf{r}}+\tilde{\mathbf{s}}_\parallel+(\tilde{d}+\tilde{t}_U)\;\hat{\mathbf{z}}}{[(\tilde{\mathbf{r}}'-\tilde{\mathbf{r}}+\tilde{\mathbf{s}}_\parallel)^2+(\tilde{d}+\tilde{t}_U)^2]^{3/2}}-
\frac{\tilde{\mathbf{r}}'-\tilde{\mathbf{r}}+\tilde{\mathbf{s}}_\parallel+(\tilde{d}+1)\;\hat{\mathbf{z}}}{[(\tilde{\mathbf{r}}'-\tilde{\mathbf{r}}+\tilde{\mathbf{s}}_\parallel)^2+(\tilde{d}+1)^2]^{3/2}}\cr
+\frac{\tilde{\mathbf{r}}'-\tilde{\mathbf{r}}+\tilde{\mathbf{s}}_\parallel+(\tilde{d}+\tilde{t}_U+1)\;\hat{\mathbf{z}}}{[(\tilde{\mathbf{r}}'-\tilde{\mathbf{r}}+\tilde{\mathbf{s}}_\parallel)^2+(\tilde{d}+\tilde{t}_U+1)^2]^{3/2}}
\Bigg)d^2 \mathbf{\tilde{r}}'\;d^2 \mathbf{\tilde{r}}+\mathbf{f}_{driving}-\tau \widetilde{m} \frac{\partial \tilde{\mathbf{s}}(t)}{\partial \tilde{t}},
}
\label{ultimate}
\end{equation}
where $\widetilde{m} =m\; \gamma^2 \alpha_L^2 \mu_0/t_L$ is the dimensionless mass, $\mathbf{f}_{driving}=\mathbf{F}_{driving}/\mu_0 M_L^2 t_L^2$ the dimensionless driving and $\tau=\zeta/\gamma\alpha_L\mu_0 M_L$ is the ration between the characteristic times of the
slider and domain dynamics.

\subsection{Equations in reciprocal space}
\label{reci}
Both the stray field and the interactions terms make (\ref{ultimateu}), (\ref{ultimatel}) and (\ref{ultimate}) non-local and practically numerically unaffordable in real space.
However, if we assume that our FFs are infinitely extended in the $xy$ plane, we can rewrite the equations of motion in reciprocal space, where the non-locality of the 
Hamiltonian disappears reducing significantly the computational cost compared to ordinary micromagnetic calculations. Naturally with the infinite extension hypothesis 
the model cannot incorporate edge effects anymore. 
Throughout the rest of the paper we will drop the tilde notation used in the previous section and any variable or coefficient is intended to be dimensionless.
Let us concentrate on the lower film equation (\ref{ultimatel}), applying a two-dimensional Fourier Transform to its stray field term we get:
\begin{equation}
\eqalign{
\mathcal{F}\Bigg[-\frac{1}{2 \pi}\int \Bigg(\frac{m_L(\mathbf{r}',t)}{\vert\mathbf{r}-\mathbf{r}'\vert}
-\frac{m_L(\mathbf{r}',t)}{\sqrt{(\mathbf{r}-\mathbf{r}')^2+1}}
\Bigg)d^2\mathbf{r}'\Bigg]\cr 
=2\pi\; m_L(\mathbf{k,t})\mathcal{F}\Bigg[-\frac{1}{2 \pi}\Bigg(\frac{1}{\vert\mathbf{r}\vert}-\frac{1}{\sqrt{\mathbf{r}^2+1}}
\Bigg)
\Bigg]
=-\frac{m_L(\mathbf{k},t)}{k}\bigg(1-e^{-k}
\bigg),
}
\end{equation}
the second step comes from the convolution theorem that, with our choice of Fourier parameter, reads $\mathcal{F}[f\ast g]=2 \pi \mathcal{F}[f] \mathcal{F}[g]$,
$m_L(\mathbf{k})$ is thus the Fourier transform of the magnetization. In the last step we made use of the Fourier transform
$\mathcal{F}[1/\sqrt{\mathbf{r}^2+a^2}]=e^{-a k}/k$ with $k=\vert \mathbf{k}\vert=\sqrt{k_x^2+k_y^2}$. Analogously, for the interaction term we get:
\begin{equation}
\eqalign{
\mathcal{F}\Bigg[-\frac{\xi}{4 \pi}\int \Bigg(-
\frac{m_U(\mathbf{r}',t)}{\sqrt{(\mathbf{r}-\mathbf{r}'-\mathbf{s}_\parallel)^2+d^2}}
-\frac{m_U(\mathbf{r}',t)}{\sqrt{(\mathbf{r}-\mathbf{r}'-\mathbf{s}_\parallel)^2+(d+t_U+1)^2}}\cr
+\frac{m_U(\mathbf{r}',t)}{\sqrt{(\mathbf{r}-\mathbf{r}'-\mathbf{s}_\parallel)^2+(d+t_U)^2}}
+\frac{m_U(\mathbf{r}',t)}{\sqrt{(\mathbf{r}-\mathbf{r}'-\mathbf{s}_\parallel)^2+(d+1)^2}}
\Bigg)d^2 \mathbf{r}'\Bigg]\cr 
=2\pi\; m_U(\mathbf{k},t)\mathcal{F}\Bigg[
-\frac{\xi}{4 \pi}\Bigg(-
\frac{1}{\sqrt{(\mathbf{r}-\mathbf{s}_\parallel)^2+d^2}}
-\frac{1}{\sqrt{(\mathbf{r}-\mathbf{s}_\parallel)^2+(d+t_U+1)^2}}\cr
+\frac{1}{\sqrt{(\mathbf{r}-\mathbf{s}_\parallel)^2+(d+t_U)^2}}
+\frac{1}{\sqrt{(\mathbf{r}-\mathbf{s}_\parallel)^2+(d+1)^2}}
\Bigg)
\Bigg]\cr
=\frac{\xi}{2}\frac{m_U(\mathbf{k},t)}{k}e^{-d \;k}e^{\mathbf{s}_\parallel \cdot \mathbf{k}}\bigg(1-e^{-k}\bigg)\bigg(1-e^{-t_U k}\bigg).
}
\end{equation}
In the last step we used the Fourier transform property $\mathcal{F}[f(\mathbf{r}+\mathbf{a})]=e^{\mathbf{a} \cdot \mathbf{k}}\mathcal{F}[f(\mathbf{r})]$.
Notice that the phase factor $e^{\mathbf{s}_\parallel \cdot \mathbf{k}}$ can be in principle moved to the magnetization $m_U$ and, going back to real space, this would lead 
to a situation in which the displacement between $O_U$ and $O_L$ is always zero, however the magnetization $m_U$ is shifted with respect to $O_U$ by a quantity $\mathbf{s}_\parallel$, see figure \ref{figure5} (c).
In this picture of a shifted magnetization, the last Fourier transform can be rewritten as:
\begin{equation}
\frac{\xi}{2}\frac{\mathcal{S}[m_U(\mathbf{k},t)]}{k}e^{-d \;k}\bigg(1-e^{-k}\bigg)\bigg(1-e^{-t_U k}\bigg),
\end{equation}
where $\mathcal{S}[f(\mathbf{k})]=e^{\mathbf{s}_\parallel \cdot \mathbf{k}}f(\mathbf{k})$ stands for the shift operation.
The transforms of the stray field and interaction terms can be now substituted into the full transform of (\ref{ultimatel}):
\begin{equation}
\eqalign{
\frac{\partial m_L}{\partial t}=\mathcal{F}\bigg[ \bigg(1-m_L(\mathbf{r},t)^2\bigg)\bigg(a_L\; m_L(\mathbf{r},t)+ h_{ext}+q_L(\mathbf{r},t)\bigg)\bigg]\cr
-\mathcal{F}\Bigg\{\bigg(1-m_L(\mathbf{r},t)^2\bigg)\mathcal{F}^{-1}\Bigg[\frac{\big(1-e^{-k}\big)}{k}\bigg( m_L(\mathbf{k},t)-\frac{\xi}{2}\mathcal{S}[m_U(\mathbf{k},t)]e^{-d \;k}\cr
\big(1-e^{-t_U k}\big)\bigg)\Bigg]\Bigg\}-\frac{b_L}{2\pi}\;k^2 m_L(\mathbf{k},t).
}
\label{fftl}
\end{equation}
The equation for the upper magnetization is obtained with the same procedure:
\begin{equation}
\eqalign{
\frac{\partial m_U}{\partial t}=\nu\mathcal{F}\bigg[ \bigg(1-m_U(\mathbf{r},t)^2\bigg)\bigg(a_U\; m_U(\mathbf{r},t)+ \frac{h_{ext}}{\xi}+q_U(\mathbf{r},t)\bigg)\bigg]\cr
-\nu\mathcal{F}\Bigg\{\bigg(1-m_U(\mathbf{r},t)^2\bigg)\mathcal{F}^{-1}\Bigg[\frac{\big(1-e^{- t_Uk}\big)}{t_U k}\bigg( \xi m_U(\mathbf{k},t)\cr
-\frac{1}{2}\mathcal{S}[m_L(\mathbf{k},t)]e^{-d \;k}\big(1-e^{-k}\big)\bigg)\Bigg]\Bigg\}-\nu\frac{b_U}{2\pi}\;k^2 m_U(\mathbf{k},t).
}
\label{fftu}
\end{equation}
From the practical point of view the transform in the first term of the two equations can be performed numerically with some Fast Fourier Transform (FFT) algorithm at each time step. 
The second term is obtained performing a FFT of the magnetization,  
constructing the new function directly in reciprocal space, coming back to real space with an inverse FFT, multiply by $1-m_L(\mathbf{r},t)^2$ and 
finally back to reciprocal space with a last direct FFT. The last term comes from the transform of the Laplacian of the magnetization and is a 
simple algebraic term. The same inverse transform method used for the non-local part of the LLGEs can be employed in the calculation of the $\mathbf{f}_U$ term in the Newton equation (\ref{ultimate}):
\begin{equation}
\eqalign{
m \frac{\partial^2 \mathbf{s}(t)}{\partial t^2}=-\frac{\xi}{2}\int  m_U(\mathbf{r},t)
\mathcal{F}^{-1}\Bigg[
\mathcal{S}[m_L(\mathbf{k},t)]e^{-d \;k}\bigg(1-e^{-k}\bigg)\cr
\bigg(1-e^{- t_Uk}\bigg)\bigg(i\frac{k_x}{k} \hat{\mathbf{i}} +i\frac{k_y}{k}\hat{\mathbf{j}}+\hat{\mathbf{k}}\bigg)
\Bigg]
d^2 \mathbf{r}+\mathbf{f}_{driving}-\tau m \frac{\partial \mathbf{s}(t)}{\partial t},
}
\label{fftn}
\end{equation}
with $i$ complex unit. Also here, after performing the FFT of the lower film magnetization, we can construct the new function in reciprocal space and apply an inverse FFT to get 
back to real space to evaluate the total force.
Applying many direct and inverse FFTs at every time step might sound computationally expensive, however being $\ell$ the size of the system, the FFT algorithm scales as $\ell \log(\ell)$, 
while a summation in real space would scale as $\ell^2$, thus performing many subsequent FFTs is still more convenient that working in real space.\\
%------------------------------------------------------
\begin{figure}
\centering
\includegraphics[height=5.0cm,angle=0]{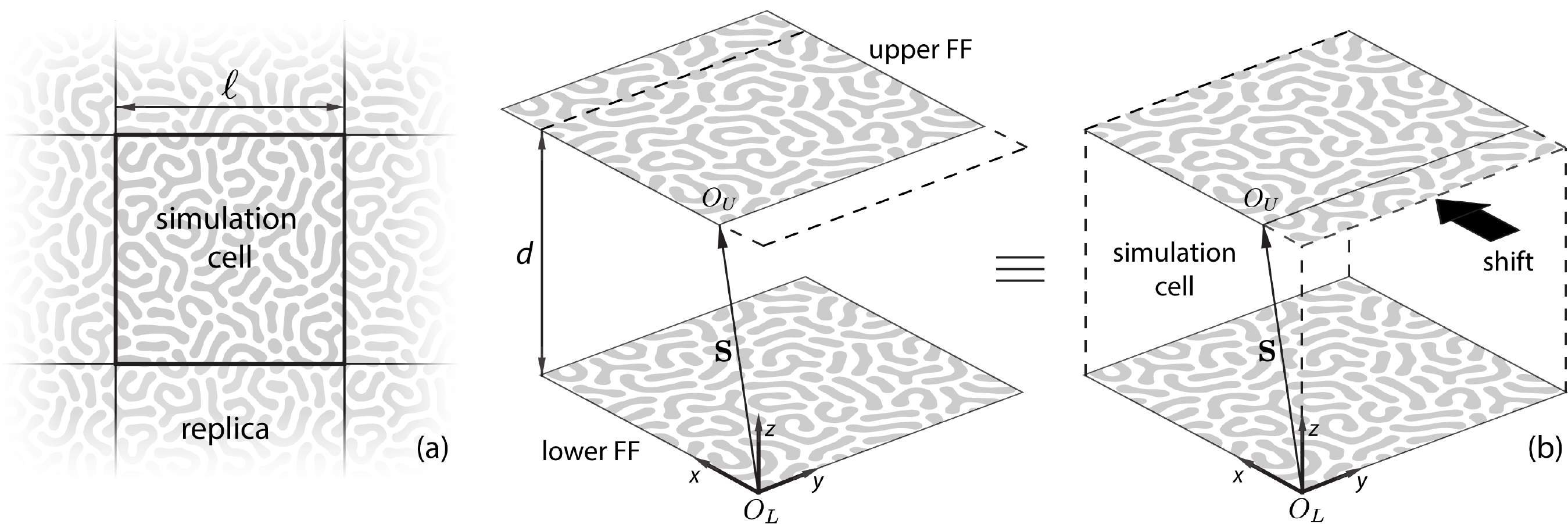}
\caption{(a) Sketch of the simulation cell with periodic boundary conditions. (b) Displacement of the upper film magnetization $m_U$ through the periodic boundary conditions.}
\label{figure6}
\end{figure}
%------------------------------------------------------
It is important to stress that using FFTs we are implicitly applying periodic boundary conditions (PBC) to our simulation cell, as illustrated in figure \ref{figure6} (a), the Fourier series of a 
function defined in the interval $\ell$ requires in fact that function to have at least periodicity $\ell$. Every magnetic dipole in the simulation cell 
feels a local field due to the rest of the infinite FF thus interacting with all the cell replicas. This is another advantage with respect to three-dimensional micromagnetic calculations for which an analytical expression 
for the stray field in reciprocal space is not available and Ewald-like summations must be performed to properly account for the long range interaction \cite{lebecki}.   
When we perform the shift operation on the magnetizations $m_U$ and $m_L$, in presence of PBC,  the portion of magnetization exiting from one mesh side is restored on the
opposite one. This is illustrated in figure \ref{figure6} (b) where the same magnetization pattern is flowing through the boundary of the simulation cell.

\subsection{Time evolution algorithm}
Equations (\ref{fftl}), (\ref{fftu}) and (\ref{fftn}) are now a set of simple differential equations in time and can be solved using finite difference methods on a squared mesh of spacing $\Delta x=\Delta y=\Delta$ with a time step $\Delta t$. 
For the time integration of the LLGEs one can use the semi-implicit first order algorithm described in reference \cite{jagla1}, the lower film equation becomes:
\begin{equation}
\eqalign{
m_L(\mathbf{k}_{ij},t+\Delta t)=m_L(\mathbf{k}_{ij},t)+\Delta t \; \Bigg\{\mathcal{F}\bigg[ \bigg(1-m_L(\mathbf{r},t)^2\bigg)\cr
\bigg(a_L\; m_L(\mathbf{r},t)+ h_{ext}+q_L(\mathbf{r},t)\bigg)\bigg]_{ij}-\mathcal{F}\Bigg\{\bigg(1-m_L(\mathbf{r},t)^2\bigg)\cr
\mathcal{F}^{-1}\Bigg[\frac{\big(1-e^{-k}\big)}{k}\bigg( m_L(\mathbf{k},t)-\frac{\xi}{2}\mathcal{S}[m_U(\mathbf{k},t)]e^{-d(t) \;k}\big(1-e^{-t_U k}\big)\bigg)\Bigg]\Bigg\}_{ij}
\Bigg\}\cr
\Bigg/\bigg(1+\Delta t\frac{b_L}{2\pi}\;\mathbf{k}_{ij}^2\bigg).
}
\label{simplettico}
\end{equation}
To calculate the magnetization $m_L(\mathbf{k}_{ij},t+\Delta t)$ in the reciprocal space mesh point $(i,j)$ one simply needs the magnetizations $m_U(\mathbf{k}_{ij},t)$ and $m_L(\mathbf{k}_{ij},t)$ in the same point at the previous time instant, and their real space counterparts
$m_U(\mathbf{r},t)$ and $m_L(\mathbf{r},t)$. Notice that the upper film magnetization which appears in the interaction term has been shifted through the PBC, for this operation we need to know $\mathbf{s}_\parallel (t)$.
An analogous equation can be obtained for the upper film:
\begin{equation}
\eqalign{
m_U(\mathbf{k}_{ij},t+\Delta t)=m_U(\mathbf{k}_{ij},t)+\nu \Delta t \; \Bigg\{\mathcal{F}\bigg[ \bigg(1-m_U(\mathbf{r},t)^2\bigg)\cr
\bigg(a_U\; m_U(\mathbf{r},t)+\frac{ h_{ext}}{\xi}+q_U(\mathbf{r},t)\bigg)\bigg]_{ij}-\mathcal{F}\Bigg\{\bigg(1-m_U(\mathbf{r},t)^2\bigg)\cr
\mathcal{F}^{-1}\Bigg[\frac{\big(1-e^{-t_U k}\big)}{t_U k}\bigg( \xi m_U(\mathbf{k},t)-\frac{1}{2}\mathcal{S}[m_L(\mathbf{k},t)]e^{-d(t) \;k}\big(1-e^{- k}\big)\bigg)\Bigg]\Bigg\}_{ij}
\Bigg\}\cr
\Bigg/\bigg(1+\nu \Delta t\frac{b_U}{2\pi}\;\mathbf{k}_{ij}^2\bigg).
}
\label{simplettico2}
\end{equation}
This equation is written in the upper film reference frame, thus, before calculating the interaction term, we have to shift the lower magnetization through the PBC, this shift is opposite to the one performed in (\ref{simplettico}).
For the Newton equation of the upper film one can use the Velocity-Verlet algorithm:
\begin{equation}
\mathbf{s}(t+\Delta t)=\mathbf{s}(t)+\Delta t\bigg(1-\frac{\Delta t}{2}\tau\bigg)\dot{\mathbf{s}}(t)+\frac{\Delta t^2}{2 m}\bigg(\mathbf{f}_U(t)+\mathbf{f}_{driving}(t)\bigg),
\label{vvp}
\end{equation}
\begin{equation}
\eqalign{
\dot{\mathbf{s}}(t+\Delta t)&=\bigg[\bigg(1-\frac{\Delta t}{2}\tau\bigg)\dot{\mathbf{s}}(t)+\frac{\Delta t}{2 m} \bigg(\mathbf{f}_U(t)+\mathbf{f}_{driving}(t)+
\mathbf{f}_U(t+\Delta t)\cr
&+\mathbf{f}_{driving}(t+\Delta t)\bigg)\bigg]/\bigg( 1+ \frac{\Delta t}{2}\tau\bigg).
}
\label{vvv}
\end{equation}
To calculate the new positions $\mathbf{s}(t+\Delta t)$ one needs to know the magnetic force $\mathbf{f}_U(t)$,
 and thus the magnetizations $m_U(\mathbf{r},t)$ and $m_L(\mathbf{k},t)$ at the previous time instant, whereas for the velocity calculation $\dot{\mathbf{s}}(t+\Delta t)$ the new forces $\mathbf{f}_U(t+\Delta t)$
are needed, this requires to solve the LLGEs to get $m_U(\mathbf{r},t+\Delta t)$ and $m_L(\mathbf{k},t+\Delta t)$. Notice that also in the calculation of $\mathbf{f}_U(t)$ we have to shift the upper film magnetization.
The two integration algorithms can thus be combined in the following way:
\begin{enumerate}
\item Having $m_U(\mathbf{r},t)$ and $m_L(\mathbf{r},t)$ the force $\mathbf{f}_U(t)$ is readily calculated; 
\item Now with $\mathbf{f}_U(t)$, $\dot{\mathbf{s}}(t)$ and $\mathbf{s}(t)$ the new upper film displacement $\mathbf{s}(t+\Delta t)$ can be evolved with (\ref{vvp});
\item With $\mathbf{s}(t+\Delta t)$ the magnetizations can be shifted and introduced in (\ref{simplettico}) and (\ref{simplettico2}) to compute $m_U(\mathbf{r},t+\Delta t)$, $m_L(\mathbf{r},t+\Delta t)$;
\item The evolved magnetizations allow to compute the new force $\mathbf{f}_U(t+\Delta t)$;
\item Having both $\mathbf{f}_U(t)$ and $\mathbf{f}_U(t+\Delta t)$ the new velocity $\dot{\mathbf{s}}(t+\Delta t)$ can be calculated with (\ref{vvv});
\item Positions, velocities and forces are updated and finally we go back to point (i).
\end{enumerate}
In this way we have a single integration of the LLGEs and of the Newton equation per time step.
$\Delta t $ must be chosen in such a way to sample with sufficient accuracy the slowest between the domain and film dynamics.
The same applies for the mesh spacing $\Delta$ which must be small enough to sample the steepest magnetization variation, i.e. smaller than the domain wall thickness. 

\section{Conclusions}
I have set up a system of equations to describe the dynamics of two bodies coated with thin ferromagnetic films, with perpendicular anisotropy, below the Curie temperature.
It is now possible to simultaneously simulate the dynamics of the two bodies, influenced by the magnetic domain interaction, and the domain 
dynamics in each film, influenced by the relative motion of the two bodies. 
Using a phase-field approach one can simulate the domain dynamics over large length scales, up to hundreds of $\mu m^2$, at a very low computational cost. The downsides of this are the absence of edge effects and the lack of generality of the model,
which applies only to perpendicular anisotropy films.\\
This new tool enables to investigate how the domain properties can influence the sliding motion of the two bodies, with potential application in the control and
actuation of micro- and nano-scale mechanical devices. On the other hand, one can also study how the body motion influences the domain properties, this can be of great interest for the design of new domain writing and domain manipulation techniques.
Finally notice that the theory developed in this paper for ferromagnetic films applies to ferroelectric films as well, allowing to evolve in time the dimensionless polarization $p(x,y)$. To this aim it is enough to substitute the magnetization $m$ with the polarization $p$, the saturation 
magnetization $M_s$ with the saturation polarization $P_s$ and the vacuum permeability $\mu_0$ with the inverse of the vacuum permittivity $1/\epsilon_0$, naturally $K_u$ and $A$ will take different values being now related to the elastic properties of the materials 
\cite{hu,sidorkin,tagantsev}.

\ack
The author is grateful to H.J. Hug, M.A. Marioni and D. Passerone from EMPA (Switzerland) for helpful discussions.
This work has been supported by grant CRSII2 136287/1 from the Swiss National Science Foundation.\\

\appendix
\section*{Appendix A}
\setcounter{section}{1}
The force $\mathbf{F}$ exerted by a magnetic field $\mathbf{B}$ on a magnetic dipole moment  $\mathbf{m}$ depends on the nature of the dipole itself \cite{boyer}. If the dipole moment is induced by
a current, the force must be calculated as:
\begin{equation}
\mathbf{F}_{c}=\nabla(\mathbf{m}\cdot \mathbf{B}),
\end{equation}
whereas, in case of a permanent dipole:
\begin{equation}
\mathbf{F}_{p}=(\mathbf{m}\cdot \nabla)\mathbf{B}.
\end{equation}
The two definitions are related by:
\begin{equation}
\mathbf{F}_{c}=\mathbf{F}_{p}+\mathbf{m}\times(\nabla\times\mathbf{B}),
\end{equation}
and they coincide if $\mathbf{B}$ is irrotational, a condition certainly valid for our field (\ref{vettoriale}).
Having a dipole moment aligned along the $z$ axis, i.e. $\mathbf{m}\equiv m\hat{\mathbf{z}}$, the expressions for $\mathbf{F}_{c}$ and  $\mathbf{F}_{p}$ simplify to:
\begin{equation}
\mathbf{F}_{c}=m \nabla Bz,
\end{equation}
\begin{equation}
\mathbf{F}_{p}=m\frac{\partial \mathbf{B}}{\partial z},
\end{equation}
and the two expressions coincide because from $\nabla\times \mathbf{B}(\mathbf{r})=0$ follows $\partial B_i/\partial r_j=\partial B_j/\partial r_i$.\\

%\References

\end{document}